\documentclass[journal=JCIM,manuscript=article]{achemso}

\usepackage{chemformula} 
\usepackage[T1]{fontenc}
\usepackage{graphicx}%
\usepackage{multirow}%
\usepackage{placeins}
\usepackage{amsmath,amssymb,amsfonts}%
\usepackage{amsthm}%
\usepackage{mathrsfs}%
\usepackage[title]{appendix}%
\usepackage{xcolor}%
\usepackage{textcomp}%
\usepackage{manyfoot}%
\usepackage{booktabs}%
\usepackage{algorithm}%
\usepackage{algorithmicx}%
\usepackage{algpseudocode}%
\usepackage{listings}%
\usepackage{subcaption}
\usepackage{ifthen}
\usepackage{hyperref}

\author{Felix Faltings}
\email{faltings@mit.edu}
\author{Hannes Stark}
\author{Tommi Jaakkola}
\author{Regina Barzilay}
\email{regina@csail.mit.edu}
\affiliation[MIT]
{CSAIL, MIT, Cambridge MA}

\title{Protein FID: Improved Evaluation of Protein Structure Generative Models}

\abbreviations{}

\begin{document}

\begin{abstract}
  Protein structure generative models have seen a recent surge of interest, but meaningfully evaluating them computationally is an active area of research. While current metrics have driven useful progress, they do not capture how well models sample the design space represented by the training data. We argue for a protein Frechet Inception Distance (FID) metric to supplement current evaluations with a measure of distributional similarity in a semantically meaningful latent space. Our FID behaves desirably under protein structure perturbations and correctly recapitulates similarities between protein samples: it correlates with optimal transport distances and recovers FoldSeek clusters and the CATH hierarchy. Evaluating current protein structure generative models with FID shows that they fall short of modeling the distribution of PDB proteins. Code is available at: \href{https://github.com/ffaltings/protfid}{https://github.com/ffaltings/protfid}.
\end{abstract}

\section{Introduction}\label{sec1}

Progress in generative protein design requires access to accurate and reliable evaluation measures. Although experimental validation remains the gold standard, in silico measures are essential to quickly develop and compare machine learning models. However, even though state-of-the-art models perform very well on current metrics, their success in practical design applications has remained relatively limited \citep{glasscock2023computational, watson2023novo,ingraham2022chroma}. For example, \cite{ingraham2022chroma} report an experimental success rate of generated structures of 3\%, while achieving much higher in silico scores. Moreover, as models continue to improve, they are beginning to outgrow current metrics, with some recent models reporting near-perfect performance \citep{campbell2024generativeflowsdiscretestatespaces}, making it difficult to compare models. Continuing advancement in this area thus requires additional, more powerful evaluation metrics to track progress.

Currently, the most commonly used in silico metrics for protein design are designability, novelty and diversity \cite{bose2024se3stochasticflowmatchingprotein, campbell2024generativeflowsdiscretestatespaces, yim2023fastproteinbackbonegeneration,yim2023se,ingraham2022chroma,watson2023novo,lin2023generating}. A structure is considered designable if there exists a sequence that folds into it. In practice, designability is evaluated by generating sequences conditioned on the generated structure and checking whether any of the sampled sequences fold back into the given structure using a folding model. On the other hand, diversity looks at how different the model generated outputs are from each other, usually assessed by looking at the number of distinct clusters over the output space. Finally, novelty checks the number of memorized samples produced by the model. 

Problematically, none of these metrics capture how well the model samples the design space represented in the training data. For example, a model could generate highly diverse, novel, and designable proteins without ever generating any beta sheets, yet beta sheets may be necessary to solve some design problems. In fact, many generative models have been observed to over-sample alpha-helices at the expense of other secondary structures \citep{lin2024out,campbell2024generativeflowsdiscretestatespaces}. Some authors address this evaluation issue by reporting the proportion of secondary structures in model-generated proteins, but the same problem persists for CATH domains \citep{orengo1997cath} or any other as-of-yet undiscovered classifications.

To address this gap, we consider using the Frechet Inception Distance (FID) introduced in \citep{heusel2017gans} and whose application to proteins was previously explored in \citep{geffnerproteina, lu2025assessing}. Building on this, we demonstrate the FID's utility as a new metric with extensive experiments across different design choices. The FID approximates the Wasserstein distance between the distribution generated by a model and a reference distribution. Hence, in contrast to designability, which evaluates \textit{individual} structures, the FID considers the entire \textit{distribution} of structures generated by the model, where a lower FID means the model captures the reference distribution better. Note that having a low FID is different from \textit{memorizing} the training distribution. A model can achieve a low FID while producing entirely novel structures. Our experiments will show that a PDB sample that has no structures in common with the reference distribution still achieves a better FID than current state-of-the-art models. 

To overcome issues with high-dimensional data, both distributions are first embedded into a latent space that captures meaningful features of the data. The FID is then computed as the Wasserstein distance between Gaussian approximations of the two distributions. To adapt the FID metric to proteins we curate a reference set of structures from the Protein Data Bank (PDB) and explore several different methods for computing embeddings based on pretrained models, including ESM3 \citep{hayes2024simulating} and GearNet \cite{zhang2022protein}.

Because it approximates the Wasserstein distance, the FID penalizes models if they miss any part of the reference distribution, thus addressing the issue raised above. We validate this in our experiments, which show that the FID can detect if models undersample FoldSeek \citep{van2024fast} or CATH clusters. Our experiments also show that the FID cannot be improved by generating invalid structures. For example, it can discriminate between PDB structures and structures folded by AlphaFold3 (AF3) \citep{abramson2024accurate} without any multiple sequence alignments (MSAs).

We then show that state-of-the-art models still have significantly higher FIDs than natural proteins. Moreover, we find that computing FIDs using different embedding dimensions and different sampling sizes leads to consistent conclusions, including the ranking between different models. This highlights the robustness of the FID as a metric for model comparison. Finally, we connect the observed gap in FIDs between natural and generated proteins to additional statistical differences. We find that the average order of inter-residue contacts in PDB structures is much higher than in generated proteins and that generated proteins display a higher redundancy of tertiary motifs (TERMs) \citep{mackenzie2016tertiary}. We additionally find that changing the generative model's sampling parameters to improve FIDs also leads to better contact orders and less TERM redundancy. 

\section{Results}

We first empirically show that generative models are saturating current metrics. Following this, we present and validate the FID as a new evaluation metric. We use it to evaluate a slate of state-of-the-art generative models, and find a gap between generated structures and natural  structures. We further illustrate the observed gap in FID by looking at other statistical differences. Finally, for some models with a tunable sampling hyperparameter, we show that tuning for FID also reduces these other statistical differences.

\subsection{Current Metrics are Starting to be Saturated}

\begin{figure}
    \centering
    \includegraphics[height=1.5in]{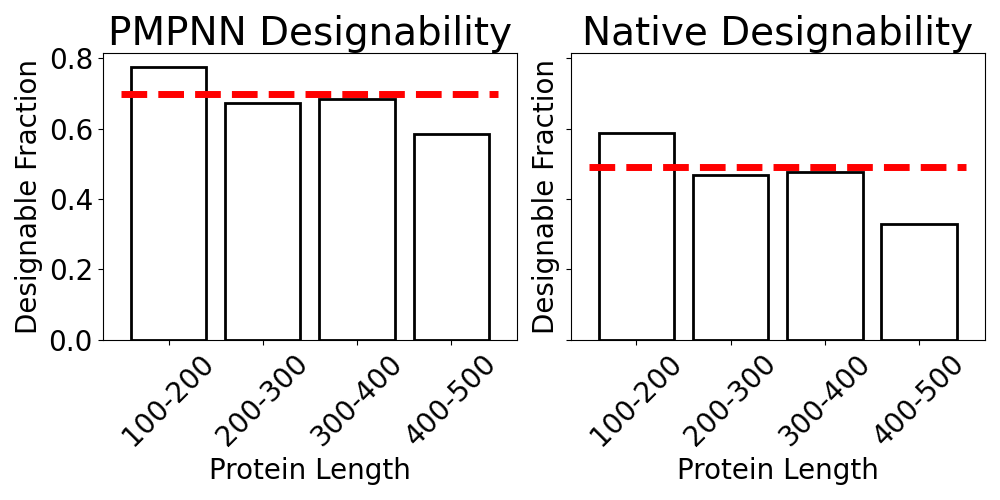}
    \caption{\textbf{Designability of PDB Proteins}. Fraction of designable structures in a curated set of PDB entries, broken down by length for sequences designed by ProteinMPNN and native sequences. The red horizontal line indicates the mean over the whole set.}
    \label{fig:pdb_designability}
\end{figure}

We show here here that recent models achieve designabilities much higher than the designability of natural proteins, making it challenging to further improve models based on designability alone. To demonstrate this, we consider single chain proteins that were released between September 2021 and March 2024 - outside of the training data of the folding model we employ, but still in distribution. Each protein is a FoldSeek cluster representative of a 100-500 length protein set. FoldSeek is a fast structural clustering algorithm designed to approximate clustering based on TMScore \cite{zhang2004scoring}, and taking cluster representatives ensures the structures are not redundant. The resulting set contains 1,024 structures.
On this set we evaluate designability by sampling 8 sequences from ProteinMPNN \citep{dauparas2022robust}, folding them with ESMFold \citep{lin2023evolutionary}, and reporting the original structure as designable if one of the 8 refolds is within 2\AA{} of the original. Fig.~\ref{fig:pdb_designability} shows the resulting designabilities broken down by length. Across all lengths, we see that a quarter of PDB structures are not considered designable. Even for shorter proteins, the designability is much lower than that achieved by generative models, and this number drops even further when considering the native protein sequences instead of those generated by ProteinMPNN. This displays how state-of-the-art protein structure generative models such as Multiflow \citep{campbell2024generativeflowsdiscretestatespaces} that reach 99\% designability for generations of similar length scales could be considered over-optimized for this metric.

\subsection{FID Metric as a Useful Tool for Evaluation}
\label{sec:fid}
A natural way to evaluate generative models is to measure the distributional discrepancy between the target distribution we want to sample from and the distribution of samples actually produced by the model. There are various ways to measure differences between probability distributions, such as the Kullback-Leibler divergence, or Wasserstein distance. However, estimating these for high dimensional distributions like protein structures is exceedingly difficult. The key idea of FID is to instead measure the difference in a lower dimensional embedding space using Gaussian approximations. Concretely, given two sets of samples, we first compute their embeddings. The distributions of the embedded samples are then approximated by Gaussians, and the 2-Wasserstein distance between the Gaussians, which can be computed with a closed-form formula, is the FID. For images, the embedding model is a pretrained image classifier \citep{heusel2017gans}, and generative models are evaluated by computing the FID to a reference set of images. For proteins, we explored several different embedding models and found ESM3 to work well. We use this in all of the following results, but investigate other embedding methods in the Appendix. As detailed in the Methods, we also found it useful to first project the ESM3 embeddings to a lower dimension. In our results we use a dimension of 32, but we sweep other dimensions in the Appendix. We find that the FID is highly robust, exhibiting consistent behavior down to as few as 4 dimensions, as demonstrated in the next section when we evaluate current generative models. Finally, we construct a reference set of structures by taking 4,991 filtered structures from the PDB, each one of which is a FoldSeek cluster representative from ESM3's training data. In addition to this, we construct a separate test set of 467 cluster representatives with no structural overlap to the reference, and whose structures were released after the ESM3 training cutoff date. See Methods for details.

\paragraph{FID detects perturbations}

\begin{figure}
\vspace{-0.9cm}
    \centering
    \includegraphics[height=1.5in]{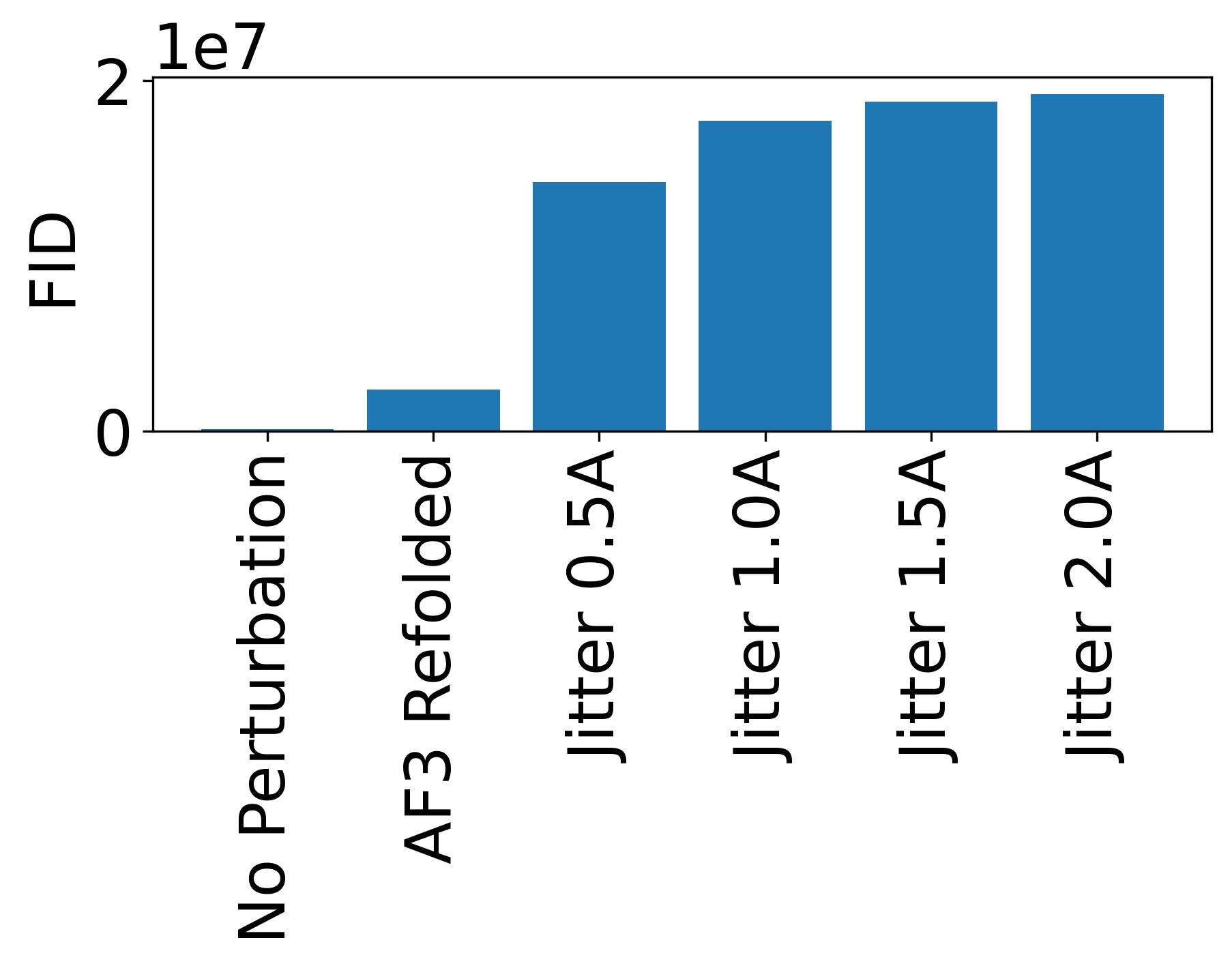}
    \caption{\textbf{FID detects perturbations}. FIDs of perturbed PDB samples. A random sample of 1,000 proteins was taken from the PDB. FIDs were computed between the reference and perturbed and unperturbed versions of the structures.}
    \label{fig:pertrubations}
\end{figure}

We first check that the FID can discriminate between physically plausible and implausible structures. As our plausible structures, we take our test set of structures. To obtain implausible structures, we apply the following perturbations to the test set,
\begin{enumerate}
    \item\textbf{Jitter}: We apply Gaussian noise to the positions of the protein's atoms.
    \item\textbf{Refolding}: We fold the proteins using AlphFold3, but without using any MSAs. These refolded structures constitute a form of perturbation, since their RMSDs to the original PDB structures are significantly worse than when folding using MSAs.
\end{enumerate}

Because the resulting perturbed structures will be out of distribution for the pretrained embedding model, their embeddings will differ from the embeddings of natural proteins and we should be able to detect this in the FIDs. The results in Fig.~\ref{fig:pertrubations} show that the FID not only detects these perturbations, but also increases monotonically with the severity of the perturbation.

\paragraph{FID recapitulates FoldSeek clustering}

\begin{figure}[th]
    \centering
    \begin{subfigure}[t]{0.5\textwidth}
        \centering
        \includegraphics[height=2.5in]{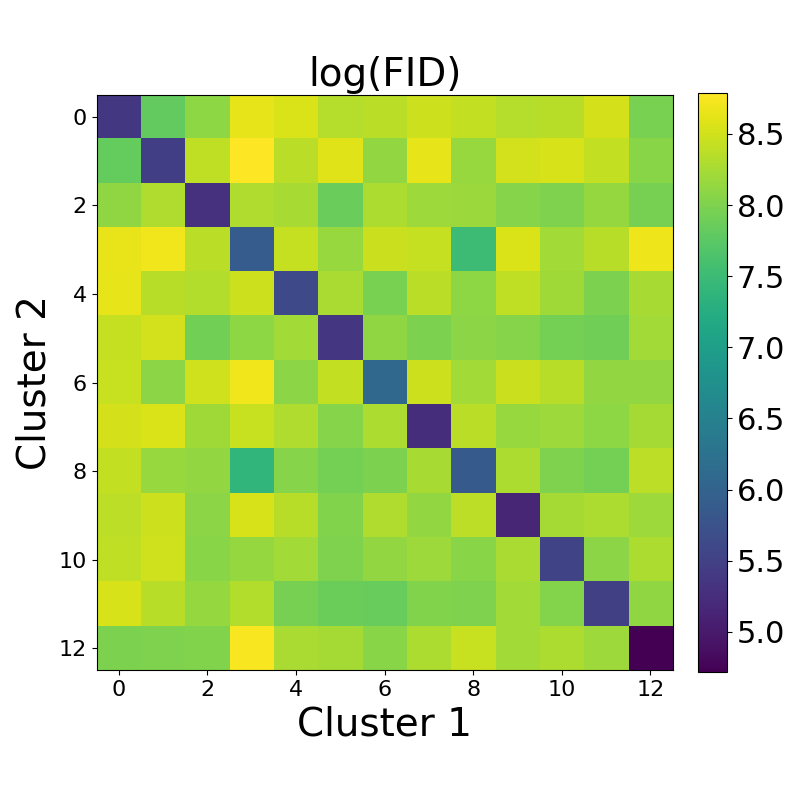}
    \end{subfigure}%
    \begin{subfigure}[t]{0.5\textwidth}
        \centering
        \includegraphics[height=2.5in]{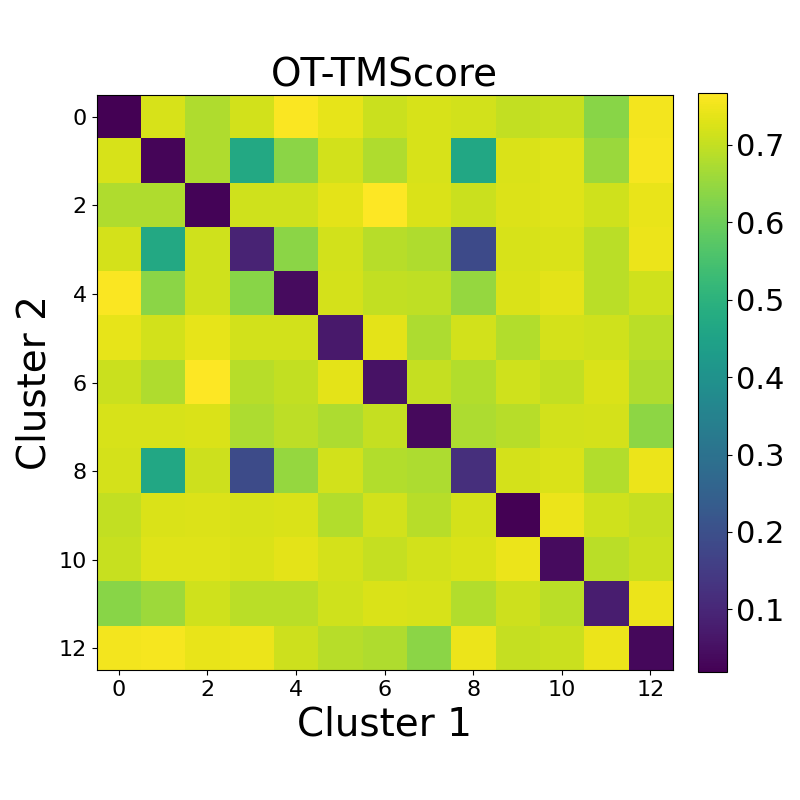}
    \end{subfigure}
    \vspace{-0.8cm}
    \caption{\textbf{FID recapitulates FoldSeek clusters and correlates with OT-TMScore.} \emph{Left:} Pairwise FIDs between FoldSeek clusters. \emph{Right:} Pairwise OT-TMScores between FoldSeek clusters. }
    \label{fig:foldseek_clusters}
    \vspace{-0.5cm}
\end{figure}

Because the FID is computed in a latent space, it is possible that the embeddings lose some structural information. We therefore verify that the FID can discriminate between more or less similar sets of structures by checking that it recapitulates the FoldSeek clustering. 

Each FoldSeek cluster represents a distribution of similar structures, where each cluster is more similar to itself than to other clusters. We check whether the FID recapitulates this by looking at the FIDs between random samples from FoldSeek clusters. For each cluster, we draw two random non-overlapping samples. We then compute FIDs between all pairs of clusters, where we always compute the FID between the first and second samples so that on the same cluster we compute the distance between two \textit{different} samples. 

We also check whether the FID can capture the \textit{degree} to which different clusters differ from each other based on the TMScore, a trusted measure of structural similarity. Because the TMScore is a metric between structures and not distributions, we compute the optimal transport distance between clusters using the TMScore as the base metric, which we call the OT-TMScore. See Methods for details. Intuitively, this metric looks at how well two samples can be soft matched based on TMScores. 

As can be seen in Fig.~\ref{fig:foldseek_clusters}, the FID between two samples from the same cluster is much lower than between two different clusters, thereby recapitulating the FoldSeek clustering. With a linear correlation of 0.76, the FID also shows good agreement with the OT-TMScore. In the figure we see for example that both metrics spot the high similarity (low FID) between clusters 8 and 3.

\paragraph{FID captures fold diversity at multiple levels}
As a distributional metric, the FID should detect whether a generative model is undersampling the variety of folds represented in the data distribution. For example, it should penalize the model for undersampling CATH clusters. To confirm this, we propose a \emph{diversity race} experiment. 

Consider a set of clustered structures. We can compute the FID of the set to a reference set after successively removing clusters in a random order. We can then compare this against removing structures at random, without regard for the clustering. Ideally, the FID should increase faster when removing entire clusters, since removing structures at random would leave most clusters represented in the remaining set. 

We can think of this as a race taking place over multiple rounds. Each racer is a set of structures. At the first round, the racers are identical. At each subsequent round, each racer drops a subset of structures. One racer will drop a cluster of structures, while the other will drop a random subset. In both cases, the FID to the reference set will increase, but they may do so at different rates.

With this analogy, we can generalize to a hierarchical clustering like CATH, where we now have more than two racers--one for each level of the hierarchy, and one baseline racer that drops structures at random. Here, we would expect the ranking of the racers to reflect the CATH hierarchy, where the higher level racer should win out, followed in order by the lower levels, and, finally, the baseline. Indeed, removing all structures from a broader category should give a higher FID than removing many narrower categories. 

Because the sizes of the clusters are not uniform, dropping different clusters will result in sets of different sizes, which may affect the FIDs. A racer should not win simply because it drops more structures in each round. To account for this, each racer always drops the same fixed number of structures in each round, but prioritizes which structures to drop based on the clustering. See Methods for a detailed description.

\begin{figure}
    \vspace{-0.4cm}
    \centering
    \begin{subfigure}[t]{0.5\textwidth}
        \centering
        \includegraphics[height=1.5in]{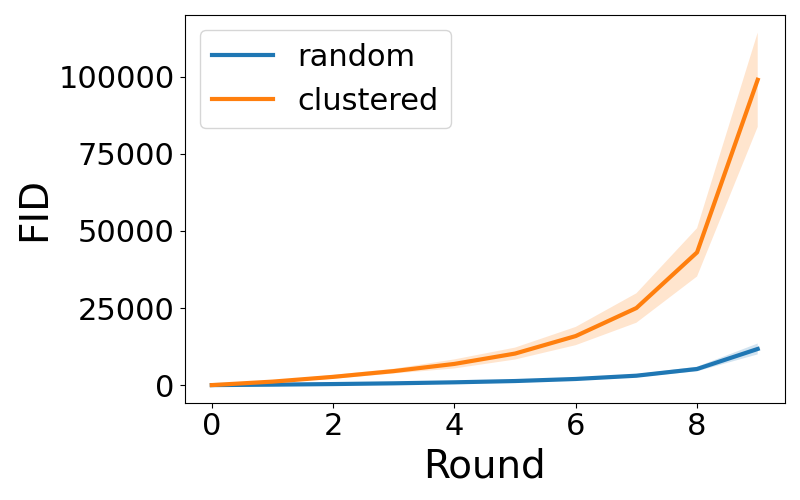}
    \end{subfigure}%
    \begin{subfigure}[t]{0.5\textwidth}
        \centering
        \includegraphics[height=1.5in]{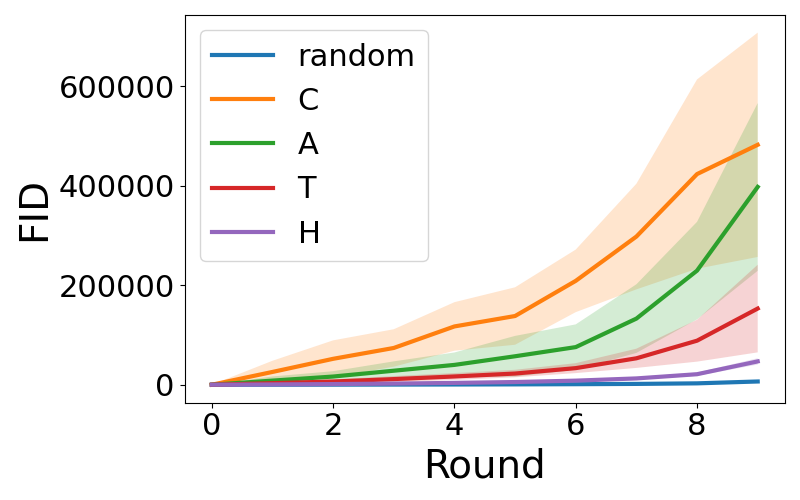}
    \end{subfigure}
    \caption{\textbf{Diversity races.} In our ``races," we increase the sizes of several ``racer" sets (starting from empty sets) by adding samples from a fixed reference set. We observe how their FIDs to the reference set drop at different speeds since some racers receive less diverse samples from clustered versions of the reference set, and since our FIDs capture meaningful structural diversity. \textit{Right}: Race conducted with CATH clusters, with one racer for each level of the hierarchy.}\label{fig:diversity-race}
    \vspace{-0.4cm}
\end{figure}

In Figure \ref{fig:diversity-race}, we carry out two such diversity races. One for a FoldSeek clustering, which results in 2 racers--the clustered order and the random order--and the hierarchical CATH clustering, which results in 5 racers--one for each of the four levels in CATH and the random order. To estimate standard deviations, we rerun the races 50 times, which allows us to ascertain that our FIDs recover both of the clusterings and meaningfully capture structural diversity. The fact that the racers in the CATH experiment are ordered according to the hierarchy of classes also suggests that the FID captures diversity at \textit{multiple} levels.

\subsection{Evaluating State-of-the-Art Generative Models}

Having validated the FID as a metric, we use it to evaluate several state-of-the-art generative models. Our results show a large remaining FID gap between generated and natural structures, which is corroborated by our observation that generated structures have lower complexity and higher substructure redundancy compared to PDB structures.

\paragraph{Current generative models are still far from natural proteins}

\begin{figure}
    \centering
    \includegraphics[height=1.5in]{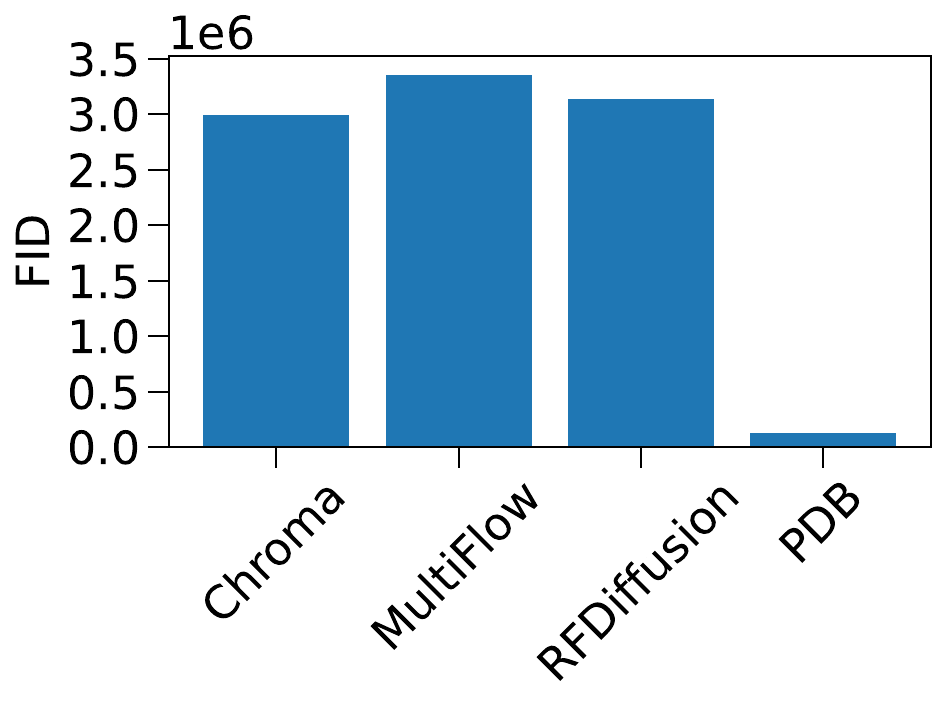}
    \caption{\textbf{FIDs of Generative Models.} FIDs achieved by state-of-the-art protein structure generative models and a set of PDB structures distinct from the reference set. The lengths of the generated proteins were chosen to match the length distribution of the PDB sample.}\label{fig:gen_model_fids}
\end{figure}

\begin{figure}
    \centering
    \includegraphics[height=1.2in]{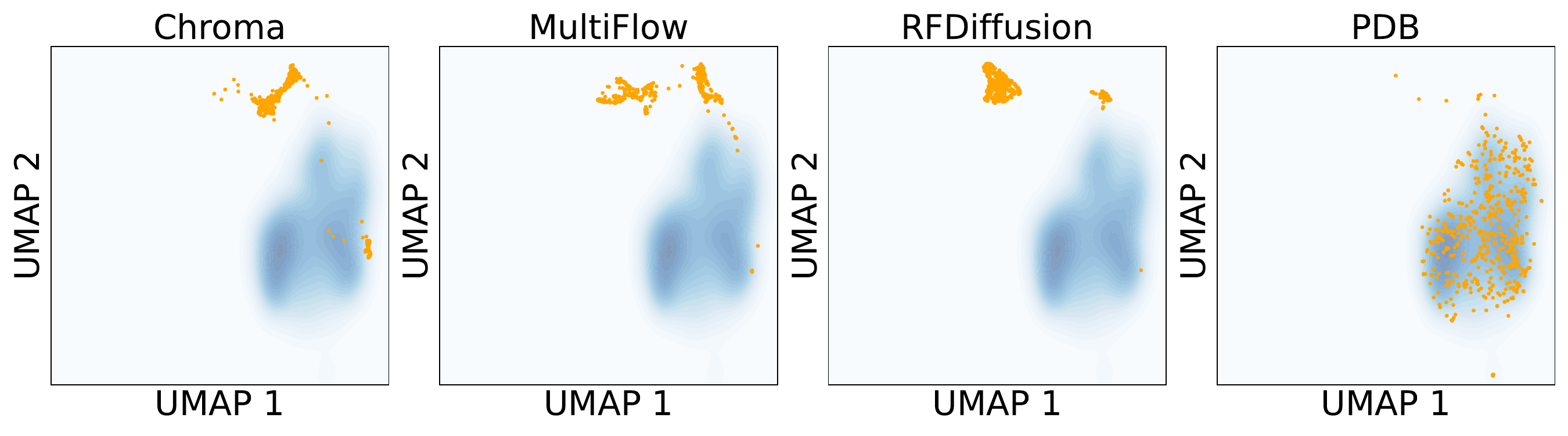}
    \caption{\textbf{Embedding UMAPs.} We visualized the UMAP embeddings of the representations for different samples of generated structures. The blue coloring shows a kernel density estimate of the reference PDB distribution. The orange points show the samples.}\label{fig:umaps}
\end{figure}

We use the FID to evaluate samples from the following generative models: MultiFlow \citep{campbell2024generativeflowsdiscretestatespaces}, RFDiffusion \citep{watson2023novo}, and Chroma \citep{ingraham2022chroma}, which we compare to our PDB test set. For each model, we generate 467 structures, to match our test set. We control for sequence length by generating one sample for each length observed in the test set, so that the resulting sets all have the same distribution of lengths. The results in Fig.~\ref{fig:gen_model_fids} show that state-of-the-art models still achieve FIDs substantially worse than natural proteins. For reference, the gap between MultiFlow and natural structures is around 1.3 times that of structures refolded by AF3 without MSAs. We also visualize the embeddings of the generated structures and the PDB structures in Fig.~\ref{fig:umaps}, where it is clear that the PDB structures better overlap with the reference set. 

\begin{figure}[ht]
    \centering
    \begin{subfigure}{0.49\textwidth}
        \centering
        \includegraphics[width=\linewidth]{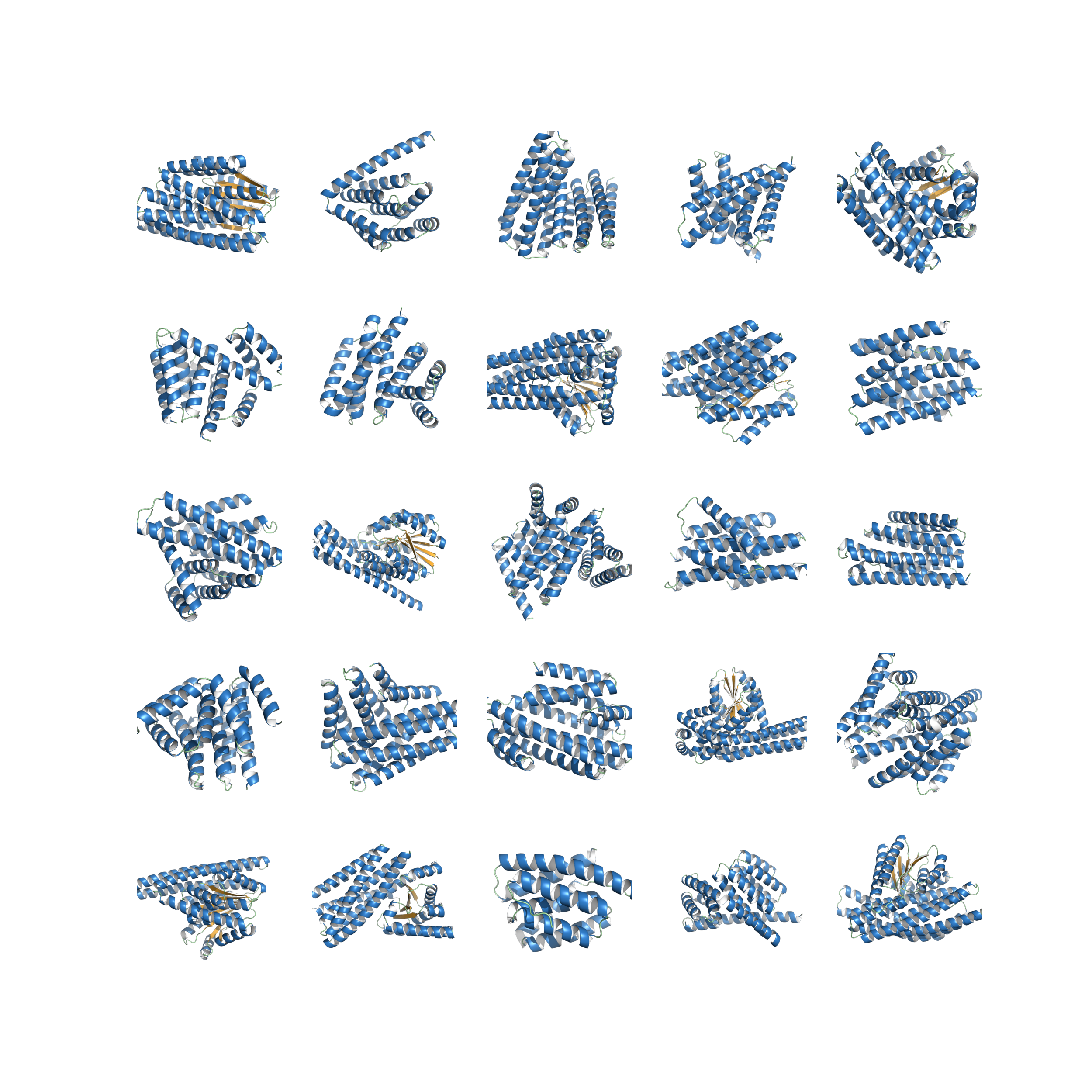}
        \caption{MultiFlow samples with negative influence on FID.}
        \label{fig:subfig1}
    \end{subfigure}
    \hfill
    \begin{subfigure}{0.49\textwidth}
        \centering
        \includegraphics[width=\linewidth]{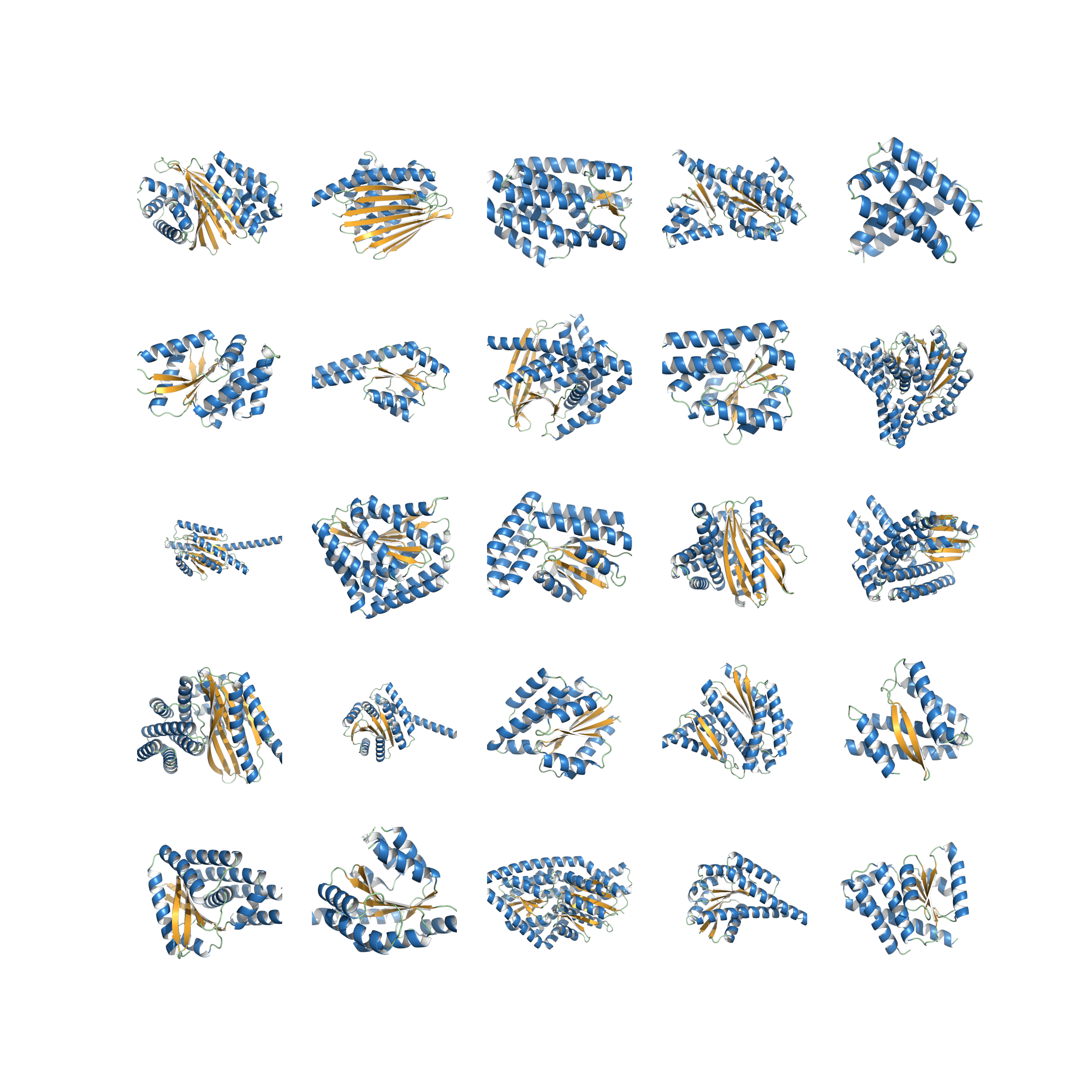}
        \caption{Multiflow samples with positive influence on FID.}
        \label{fig:subfig2}
    \end{subfigure}
    \caption{\textbf{FID prefers structures with diverse motifs.} Samples whose weights had a positive gradient (left) and negative gradient (right) for the FID. A positive gradient indicates samples that have a negative effect on the FID and vice versa for negative gradients.}
    \label{fig:mf_fid_optimizers}
\end{figure}

Finally, in order to better illustrate the gap between generated structures and natural structures, we estimate which samples improve or worsen the FID to the reference set the most. We first assign weights to each generated sample, indicating how much each sample contributes to the FID. For example, assigning a weight of 0 is the same as removing a sample entirely. We then consider the derivative of the FID with respect to these weights and visualize which samples have a positive or negative influence, corresponding respectively to negative and positive derivatives. See Methods for more details. We see in Fig.~\ref{fig:mf_fid_optimizers} that for MultiFlow, the samples with a negative influence all consist of large alpha-helical bundles, whereas the positive influence samples are more diverse, also confirming our previous observation that the generative models tend to under-sample beta sheets. However, we note that the FID is not only capturing a difference in secondary structure, since some of the negative samples also contain beta sheets.

\paragraph{FID is Robust to Sample Sizes and Embedding Dimensions}

\begin{figure}
    \centering
    \begin{subfigure}[b]{0.95\textwidth}
        \centering
        \includegraphics[width=\textwidth]{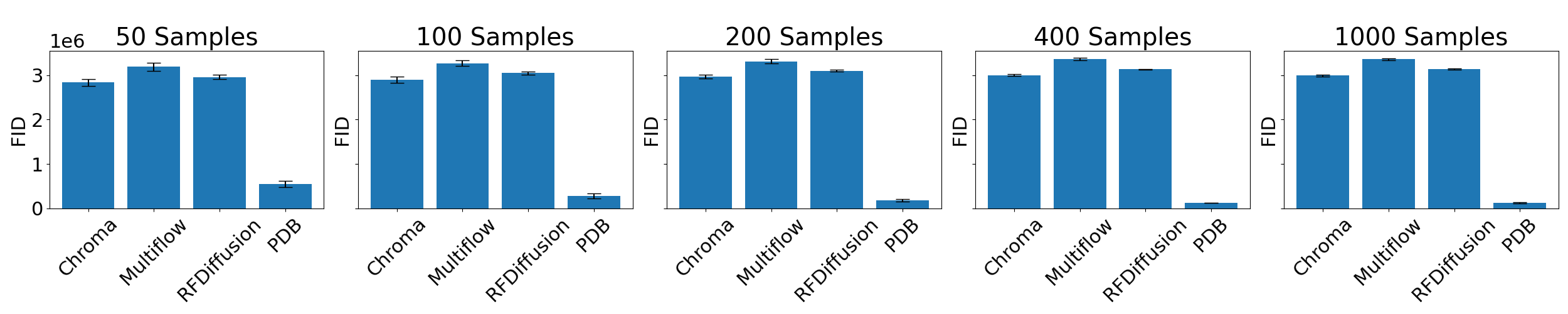}
    \end{subfigure}
    
    \vspace{1em}

    \begin{subfigure}[b]{0.95\textwidth}
        \centering
        \includegraphics[width=\textwidth]{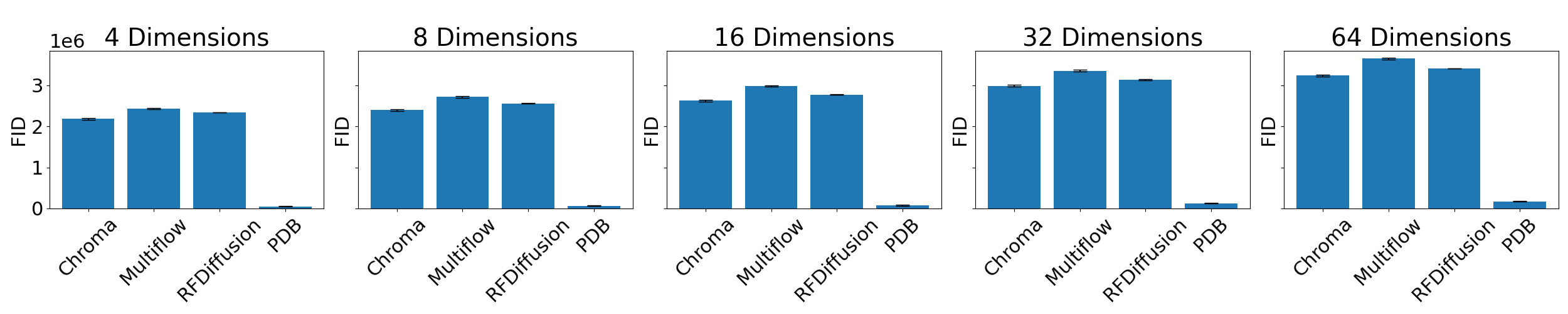}
    \end{subfigure}
    
    \caption{\textbf{Robustness Analysis} \textit{Top}: FIDs of generative models for different numbers of samples. \textit{Bottom}: FIDs of generative models for different dimensions of embeddings. In all cases, we see that the order of the models is preserved.}
    \label{fig:robustness}
\end{figure}
As an evaluation measure, it is important that the FID gives consistent results. We thus recompute the FIDs of the generative models above using varying sample sizes and varying embedding dimensions as shown in Fig.~\ref{fig:robustness}. We see that even with a sample size as small as 50 structures, the ordering of the models is preserved and the variances are relatively small. Similarly, when projecting to as low as 4 PCA dimensions, we still observe a consistent ranking between models. This demonstrates the robustness of the FID as an evaluation metric.

\paragraph{Generative models generate proteins with lower complexity than natural proteins}
Another possible explanation for the gap in FIDs between generated and natural proteins is that natural proteins may be more complex. As a proxy to address the complexity of protein structures, we count the number of higher order contacts in the structure. More specifically, we consider potential contacts (PCs) as defined in \citep{mackenzie2016tertiary}, where two backbone positions form a PC if their side chains are close to each other when considering all combinations of rotamers and side chains at both positions. We then define the order of a contact as the number of distinct secondary structures that appear between the contact positions. For example, a contact that occurs within the same secondary structure, such as in an alpha-helix, would have an order of 0. The average and median contact orders for samples from the PDB and generative models are reported in Table~\ref{tab:complexity_results}. We see that on average, the PDB contains structures with significantly higher order contacts than generated proteins.

\begin{table}
\centering
\begin{tabular}{l r r}
\toprule
& \multicolumn{2}{c}{\textbf{Contact Order}} \\ \cmidrule{2-3}
\textbf{Set} & \textbf{Mean} & \textbf{Median} \\
\midrule
multiflow & 3.3 & 2 \\
chroma & 4.0 & 2 \\
rfdiffusion & 1.9 & 0 \\
pdb & 7.6 & 3 \\
\bottomrule
\end{tabular}
\captionsetup{width=\textwidth}
\caption{\textbf{Natural structures have more high order contacts than generated structures.} Average and median order of contacts across all proteins in a random sample from the PDB and different generative models. A contact is a pair of backbone positions that could potentially form a contact. The order of the contact is the number of secondary structures between the pairs of positions. Contacts in the same secondary structure have order 0.}
\label{tab:complexity_results}
\vspace{-0.4cm}
\end{table}

\begin{figure}
    \centering
    \includegraphics[height=1.8in]{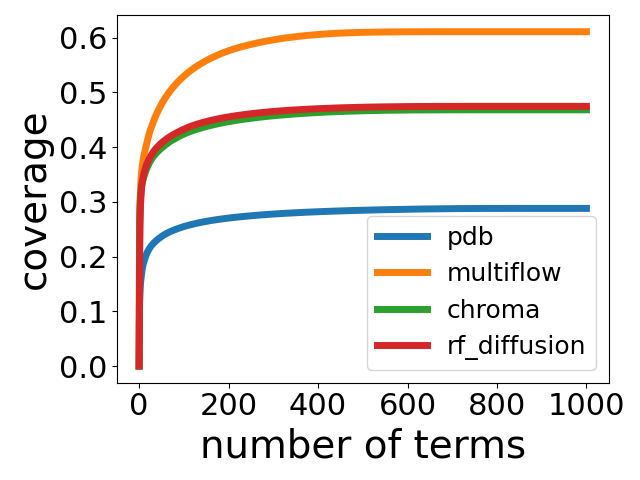}
    \caption{\textbf{TERM coverage.} Number of TERMs from \citep{mackenzie2016tertiary} needed to cover samples from the PDB and generative models.}
    \vspace{-0.4cm}
    \label{fig:terms}
\end{figure}

\paragraph{Generative models generate proteins with low substructure diversity}
To explain the gap in FIDs between natural and generated proteins, we compare their diversity of substructures. Specifically, we consider the number of tertiary structural motifs (TERMs) from \citep{mackenzie2016tertiary} needed to explain each set of structures. A TERM is a small neighborhood around an amino acid, including potentially contacting amino acids. A structure can then be seen as a set of backbone atom positions and potential contacts (PCs) as defined above. A TERM is said to cover a set of positions and PCs if it can be aligned to those positions with a low enough RMSD. A structure is said to be explained by a set of TERMs if they cover all its backbone positions and PCs \citep{mackenzie2016tertiary}. To assess the substructure diversity of a set of structures, we take the top 1000 TERMs from the original paper \citep{mackenzie2016tertiary}. We then look at how much of the structures can be explained using those TERMs. As can be seen in Fig.~\ref{fig:terms}, the generated proteins can be explained with much fewer TERMs than the PDB sample, indicating they contain a lower diversity of substructural motifs.

\paragraph{Improving FID can improve substrcture diversity and complexity}

\begin{figure}[ht]
    \centering
    \begin{subfigure}[b]{0.3\textwidth}
        \centering
        \includegraphics[width=\textwidth]{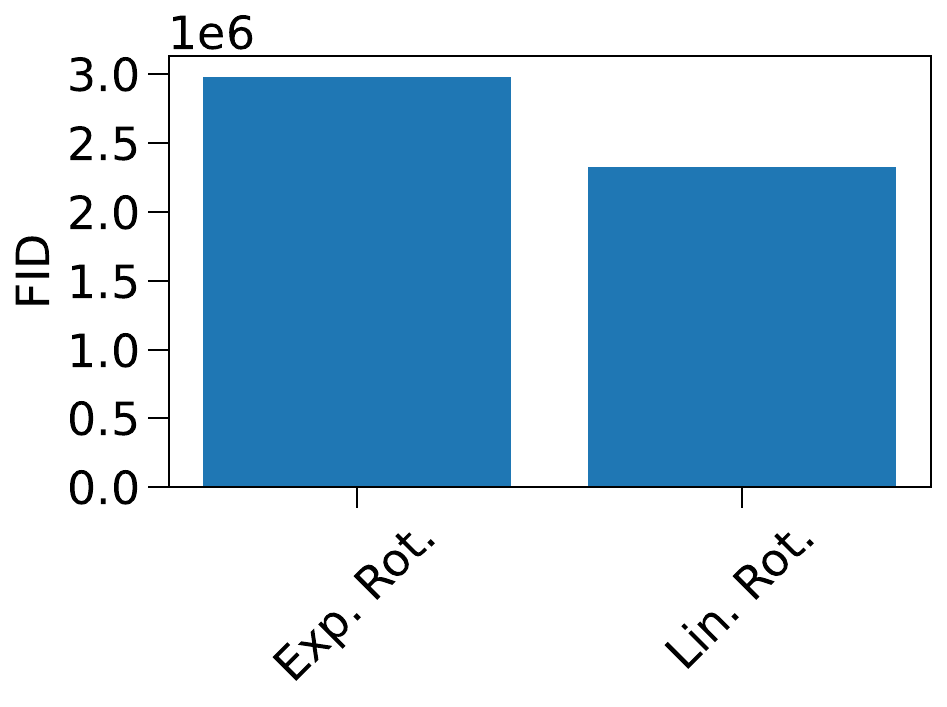}
        \caption{}
    \end{subfigure}
    \hfill
    \begin{subfigure}[b]{0.3\textwidth}
        \centering
        \includegraphics[width=\textwidth]{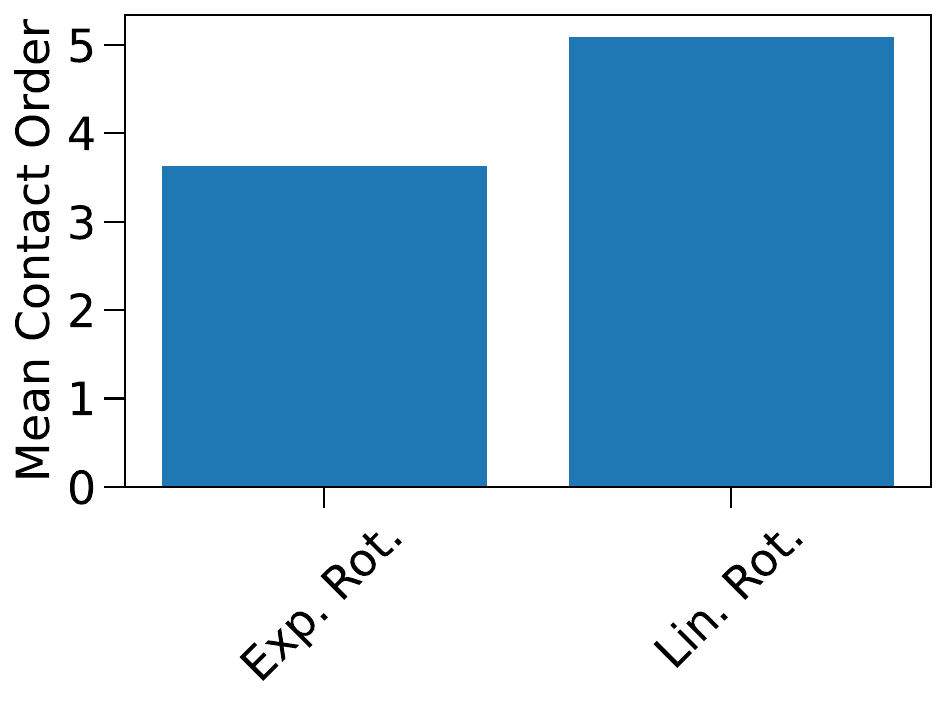}
        \caption{}
    \end{subfigure}
    \hfill
    \begin{subfigure}[b]{0.3\textwidth}
        \centering
        \includegraphics[width=\textwidth]{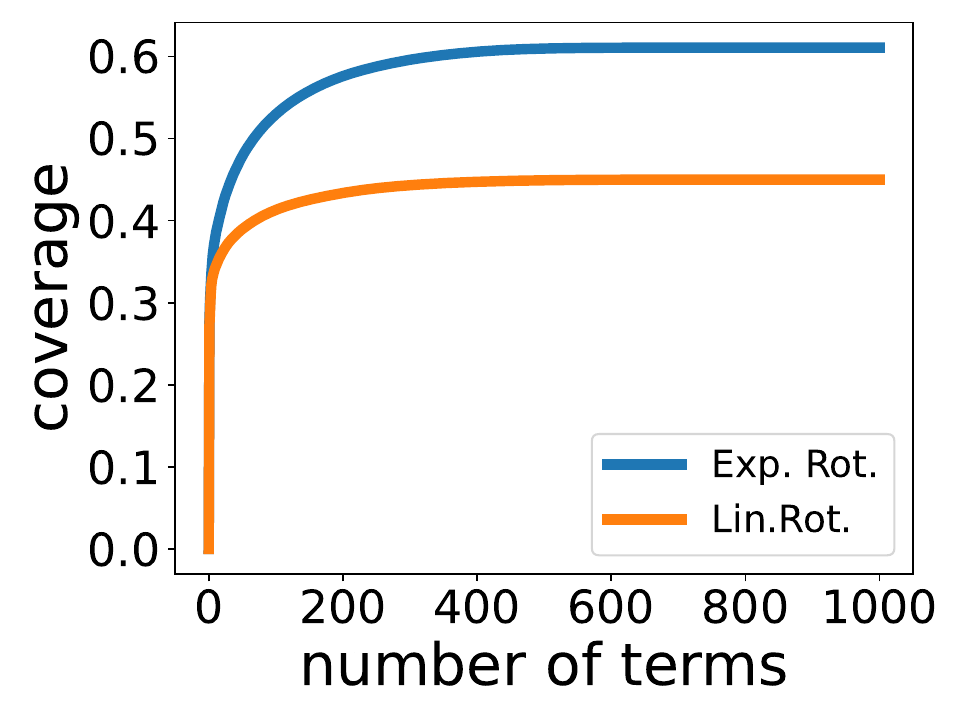}
        \caption{}
    \end{subfigure}
    \caption{\textbf{Effect of MultiFlow Sampling Parameters}. We compare samples from MultiFlow using the exponential schedule for rotations used in the original paper, versus the linear schedule the model was trained with. We report the FID (a), average contact order (b) and TERM coverage (c). The linear schedule leads to better results across all three metrics.}
    \label{fig:mf_linrots}
\end{figure}

\begin{figure}
    \vspace{-0.4cm}
    \centering
    \includegraphics[height=1.5in]{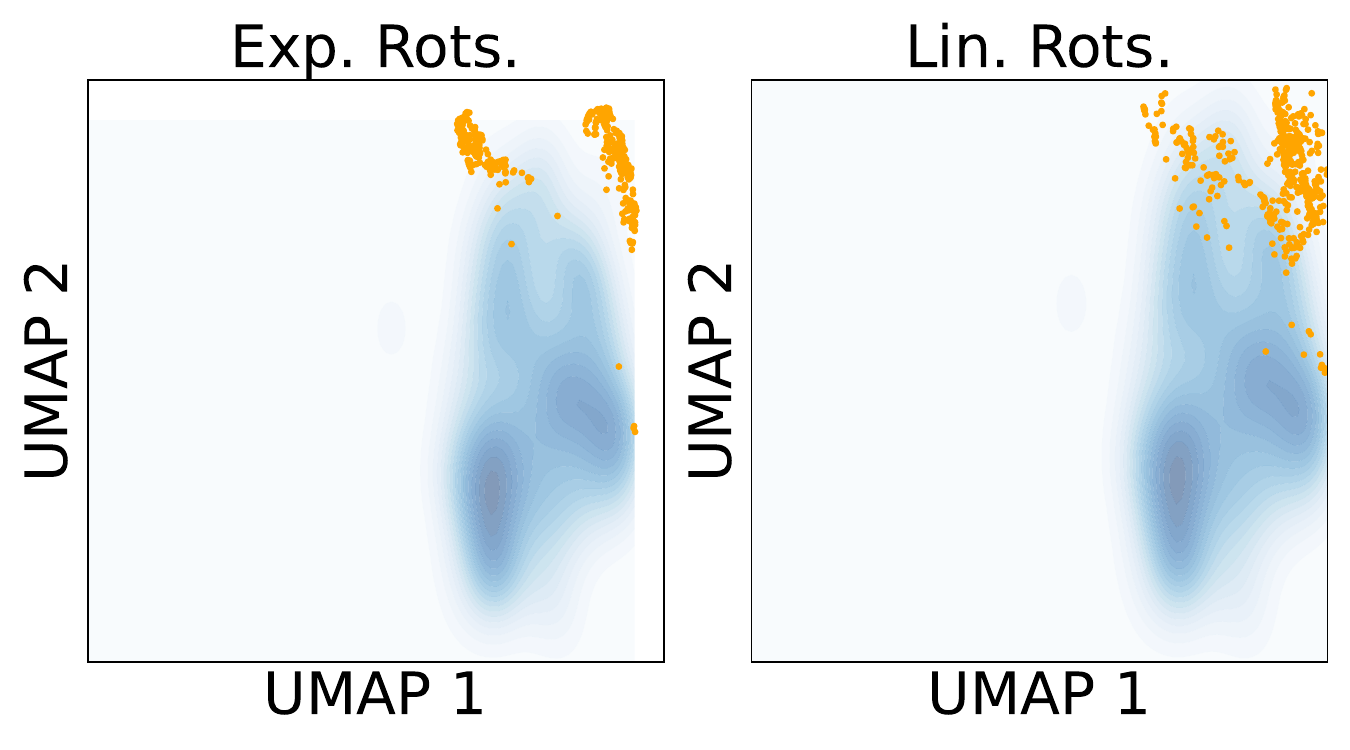}
    \caption{\textbf{Changing sampling hyperparameters increases overlap of samples with reference.} UMAP embeddings of proteins generated by MultiFlow when using the exponential rotation schedule from the paper versus the linear rotation schedule used during training. We see better overlap between model generated samples and reference clusters when using the linear schedule.}
    \label{fig:lin_rot_umaps}
\end{figure}

Frame-based flow models \cite{campbell2024generativeflowsdiscretestatespaces, bose2023se} have previously found that using different hyper-parameters at sampling time than at training time leads to better designability. In particular, the noise schedule of the frame rotations is annealed much faster than the translations. Here we show that, conversely, using the original noise schedule improves FIDs. We sample from MultiFlow either with the hyperparameters from the paper (Exponential), or the hyperparameters used during training (Linear), and find that the training hyperparameters lead to better FIDs (Fig.~\ref{fig:mf_linrots} (a)), higher average contact orders (Fig.~\ref{fig:mf_linrots} (b)), and more substructural diversity (Fig.~\ref{fig:mf_linrots} (c)), but lower designability. We also see improved coverage in the UMAP projections (Fig.~\ref{fig:lin_rot_umaps}). Thus, we see that changing the sampling schedule to improve FIDs also improves the substructure diversity and complexity of generated proteins.

\FloatBarrier

\section{Discussion}
In this work, we argue that current evaluation metrics of generative protein structure models insufficiently evaluate how well generated structures represent the training data. We thus propose to use the FID, which has proven successful in computer vision. We show that the FID can penalize models for lack of diversity, or for missing parts of the training distribution, and discriminates between physically plausible and implausible structures. Novel structures also still achieve good FIDs, showing that the FID does not simply reward memorization of training samples. We then re-evaluate state-of-the-art generative models in terms of FIDs and find that the generated structures are still far from natural structures, even though they achieve near-perfect designability. We also find that these results are robust, giving consistent results for different sample sizes or embedding dimensions used to compute the FIDs. Moreover, we find that we can improve the FID of some of these models by changing their sampling hyperparameters, leading to a corresponding improvement in substructure diversity and structure complexity. 

One limitation that is not fully explored in this work is the embeddings used to compute the FID. While we use pretrained embedding models, there may be better ways to learn these embeddings specifically for the purpose of computing FIDs. These embeddings could then also be tuned towards specific types of errors that protein design practitioners care about.

In conclusion, the FID appears to be a viable evaluation measure, and a strong complement to existing measures like designability. This additional metric is all the more pertinent as current metrics begin to saturate. Indeed, we found that recent models achieve designabilities close to 99\%, whereas natural proteins in the PDB of a similar length are only around 80\% designable. This makes it unclear whether further optimizing designability will lead to meaningful progress. Reducing the FID gap between natural and generated proteins should thus serve as a new target for improving generative models.

\section{Methods}

Here we present all the computational methods used to produce our results.

\paragraph{FID Formula} Given two samples of structures $X = \{x_1,\dots,x_N\}$ and $Y = \{y_1,\dots,y_M\}$, we first compute embeddings $E(x_i)\in\mathbb{R}^p,$ and $E(y_j)\in\mathbb{R}^p$, of dimension $p$. Let $E_X$ denote the $p\times N$ matrix of embeddings $[E(x_1),\dots,E(x_N)]\in\mathbb{R}^{p\times N}$, and similarly for $E_Y$. We approximate the distributions of the embeddings using Gaussians. This amounts to estimating parameters,
\begin{equation}
    \mu_X = \frac{1}{N}\sum_{i=1}^N E(x_i),\quad \Sigma_X = \frac{1}{N}\sum_{i=1}^N(E(x_i) - \mu_X)(E(x_i) - \mu_X)^T,
\end{equation}
with $\mu_Y$ and $\Sigma_Y$ defined similarly. The FID is then the Wasserstein distance between these Gaussian approximations, which is given by the formula,
\begin{equation}
\mathcal{W}_2(\mu_X,\Sigma_X,\mu_Y,\Sigma_Y) = \|\mu_X - \mu_Y\|^2_2 + \text{Tr}(\Sigma_X + \Sigma_Y - 2(\Sigma_Y\Sigma_X)^{1/2}).
\end{equation}

\paragraph{Embeddings} In our experiments, we use the embeddings from ESM3. We mask out the protein sequences and only use the structure tokens so the embeddings only encode the structure. Since the ESM3 structure tokens only depend on backbone frames, we can evaluate any generative model that generates the full backbone. This includes models like Chroma and RFDiffusion, but excludes models like Genie2 which only generate $C_\alpha$ atoms. We average the amino acid embeddings produced by ESM3 to get a protein embedding of dimension 1,536. However, we found that the covariance matrices of the embeddings were nearly degenerate, with only a few dominant principal components (see Fig.~\ref{fig:cov_ranks} in the appendix). This lead to numerical issues when computing the matrix square roots in the FID formula. We thus found it beneficial to project onto the first 32 principal components, giving embeddings of dimension 32. We compute the PCA using the embeddings of both the reference set and the evaluation set in order to emphasize differences between the two. 

\paragraph{Reference Set} In order to construct our reference set, we start from the filtered training set of OpenFold \citep{Ahdritz2022.11.20.517210}, filtered for structures of 100 to 500 residues obtained either via X-ray diffraction or electron microscopy with a resolution under 3\AA. We cluster the resulting set using FoldSeek, and only keep structures with a single identified protein chain, since all the generative models we consider only generate single-chain structures. We cluster the resulting set using FoldSeek and split the clusters based on ESM3's training date cutoff of 12/01/2020, keeping all clusters before the cutoff for our reference set, and clusters after the cutoff for our test set. The release date of a cluster is taken as the earliest release date of any of its members. Finally, we sample a representative from each cluster to give a representative, diverse, and non-redundant set of structures. This results in 4,991 reference structures and 467 test structures. We verify that the test set does not have significant structural overlap with the reference by computing the highest TMScore of each structure in the test set to any structure in the reference using the Foldseek easy-search software \cite{van2024fast}. The resulting mean TMScore of 0.24 indicates there is no significant overlap. 

\paragraph{Optimal Transport TM Score}
The TMScore is a similarity metric between two proteins structures. Here we describe how to turn it into a metric between sets of structures. Given two sets of proteins $P_1,\dots,P_N$ and $Q_1,\dots,Q_M$, we can compute a matrix of pairwise TMScores,
\begin{equation}
    D_{ij} = \text{TMScore}(P_i, Q_j).
\end{equation}
We turn this into a symmetric\footnote{The TMScore is not symmetric} cost matrix,
\begin{equation}
    C_{ij} = 1 - (D_{ij} + D_{ji})/2
\end{equation}
The OT-TMScore between the two samples is the solution to the optimal transport problem,
$$
\begin{aligned}
    & \underset{P_{ij}}{\min} && \sum_{ij}P_{ij}C_{ij} \\
    & \text{subject to} \\
    & && \sum_jP_{ij} = 1\\
    & && \sum_iP_{ij} = 1 \\
    & && P_{ij} \geq 0 \\
\end{aligned}
$$
Informally, this metric looks at how well the structures in each sample can be (soft) matched to each other based on the TMScore.

\paragraph{Diversity Race} Here we describe the details of the diversity race in Section~\ref{sec:fid}. The race proceeds over $L$ rounds. Each racer, $r_1,\dots,r_M$ is a set of structures. Let $r_i^j$ denote racer $i$ at round $j$. The sets are decreasing in size, $r_i^1 \supset r_i^{2} \supset\dots\supset r_i^L$, and each racer starts off identical $r_i^1 = r$, where $r$ contains $N$ structures. At each round, we compute the FID to a reference set $R$, $s_i^l = \text{FID}(r_i^l, R)$.

Each racer differs in which structures are dropped at each round. Consider racer $i$, which prioritizes dropping structures according to clusters, $c_1,\dots,c_K$, where $\cup_{i=1}^Kc_i = r$. We order the $N$ structures in $r$ according to the clustering. If $n_1,\dots,n_K$ are the sizes of the clusters, then the first $n_1$ structures are from cluster $c_1$, the next $n_2$ from $c_2$, and so on. The racer then drops structures according to this ordering, thus prioritizing dropping a whole cluster before moving on to the next one.

\paragraph{FID Optimization}
In order to probe the influence of each sample on the FID, we compute the FID using a weighted sample and consider the derivative of the FID with respect to the weights. Samples with a negative impact on the FID should receive positive gradients (increasing the FID is worse). We largely follow the methodology in \cite{kynkaanniemi2022role}. Given a sample of structures $x_i$, we compute embeddings $E(x_i)$. We then assign weights $w_i$, and compute the weighted mean and covariance,
\begin{equation}
    \mu_X(w) = \sum_{i=1}^Nw_iE(x_i),\quad\Sigma_X(w) = \sum_{i=1}^Nw_i(E(x_i) - \mu_X)(E(x_i) - \mu_X)^T, 
\end{equation}

where we parameterize the weights as $w_i = \frac{1}{\sum_{i=1}^N\exp(u_i)}\exp(u_i)$ so that they are nonnegative and sum to 1. We then compute the FID as before using a differentiable implementation and compute the derivatives when $w_i=1,\,\forall i$. When visualizing negative and positive samples, we first deduplicate the structures to avoid redundancy since similar structures will have a similar influence on the FID. Concretely we build a graph connecting any two structures with an embedding distance higher than a threshold and greedily approximate a maximum independent set in the graph. The threshold is set as the median of all pairwise embeddings distances.

\paragraph{TERM Analyses}

TERM analyses are performed using the MASTER \cite{zhou2020c++} and ConFind \cite{holland2018contact} software packages to respectively search for TERM matches and identify potential contacts. To compute the coverage of a target set of structures, we first identify all potential contacts and residues, forming a set of elements $U$. We then take the 1,000 first TERMs from \cite{mackenzie2016tertiary} to use as query TERMs. Using MASTER, each query is matched to TERMs in the target set. Potential contacts and residues in the matched target TERMs are considerd covered by the query. To estimate how to cover $U$ with the least number of query TERMs, we greedily select TERMs that cover the most remaining elements in $U$, as done in \cite{mackenzie2016tertiary}.

\bigskip

\FloatBarrier

\bibliography{iclr2025_conference}

\newpage

\begin{appendices}

\section{Comparison of Embedding Methods}\label{secA1}

Here we explore other ways of computing the embeddings. First, we consider using GearNet \cite{zhang2022protein} as an alternative to ESM3. GearNet is a contrastively pretrained structure embedding model. Similarly to ESM3, its learned embeddings should capture features of protein structures.

We first illustrate that for both ESM3 and GearNet, the embedded distributions have extremely rank-deficient covariance matrices, motivating the need to project the embeddings to a lower dimension. We then present results for all the same experiments as in the main text, but with ESM3 embeddings replaced by GearNet embeddings. We will show that the FIDs computed using GearNet embeddings do not behave as cleanly as those computed using ESM3 embeddings. Next, we consider using the token embeddings from ESM3 instead of using the average embedding and see that the resulting FID focuses more on local rather than global structure.

\paragraph{Degeneracy of Embeddings}

\begin{figure}[th]
    \centering
    \begin{subfigure}[t]{0.5\textwidth}
        \centering
        \includegraphics[height=1.5in]{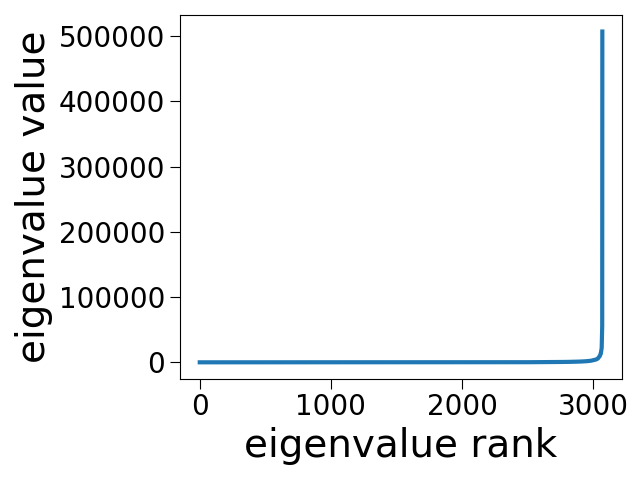}
    \end{subfigure}%
    \begin{subfigure}[t]{0.5\textwidth}
        \centering
        \includegraphics[height=1.5in]{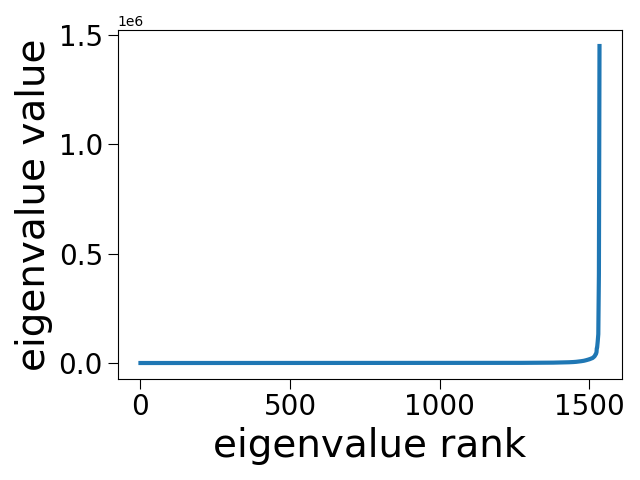}
    \end{subfigure}
    \caption{\textbf{Degeneracy of Embeddings} Sorted eigenvalues of covariance matrix of embedded distribution. \textit{Left}: Embeddings from GearNet. \textit{Right}: Embeddings from ESM3.}
    \label{fig:cov_ranks}
\end{figure}

For both GearNet and ESM3 we compute the embeddings of our reference set. We then compute the eigenvalues of the covariance matrix of the resulting embeddings, which we plot in sorted order in Fig.~\ref{fig:cov_ranks}. In both cases we can see from the small number of dominating eigenvalues that the embedded distributions are spread over a very low-dimensional subspace. We believe this to be the cause of numerical instabilities we observed when trying to compute FIDs using the full dimensional embeddings. This motivated our approach of first projecting the embeddings to a lower dimensional subspace.

\paragraph{GearNet Embeddings}

\begin{figure}
\vspace{-0.9cm}
    \centering
    \includegraphics[width=.48\textwidth]{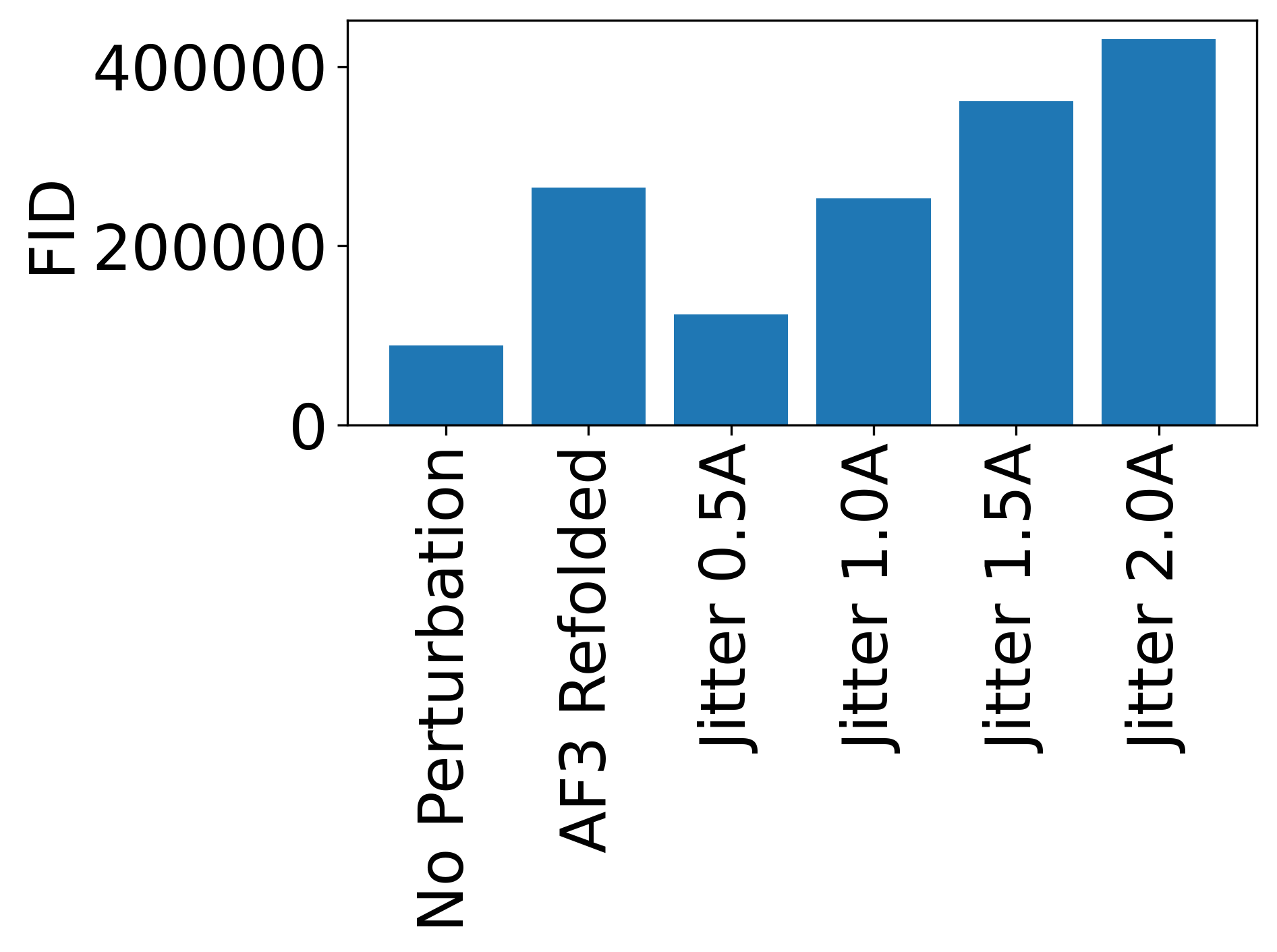} 
    \vspace{0.5cm} 
    \includegraphics[width=.48\textwidth]{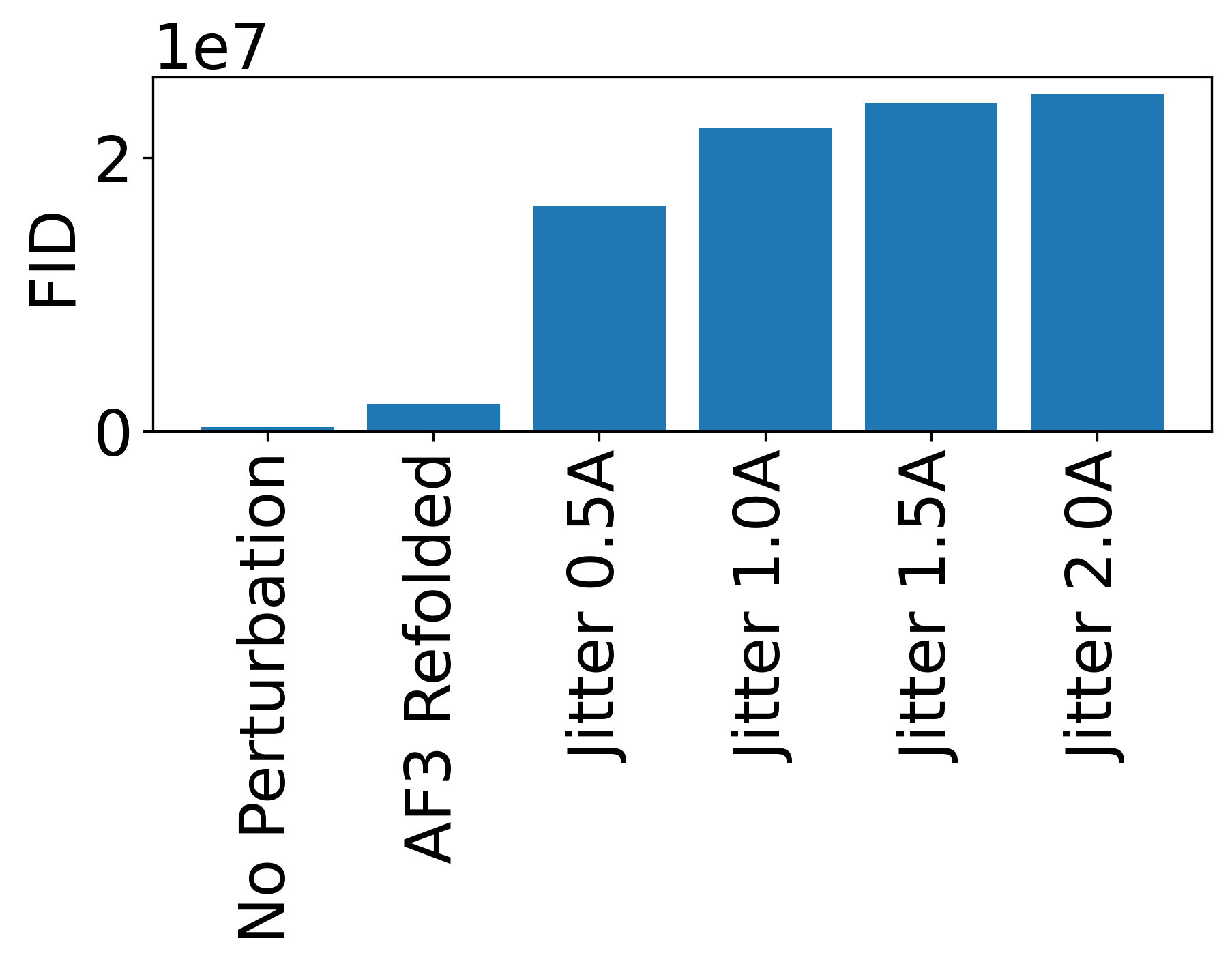}
    \caption{\textbf{FID Ladders}. FIDs of perturbed PDB samples for GearNet (left) and split ESM3 embeddings (right).}
    \label{fig:app_perturbations}
\end{figure}

\begin{figure}[th]
    \centering
    \begin{subfigure}[t]{0.5\textwidth}
        \centering
        \includegraphics[height=2.5in]{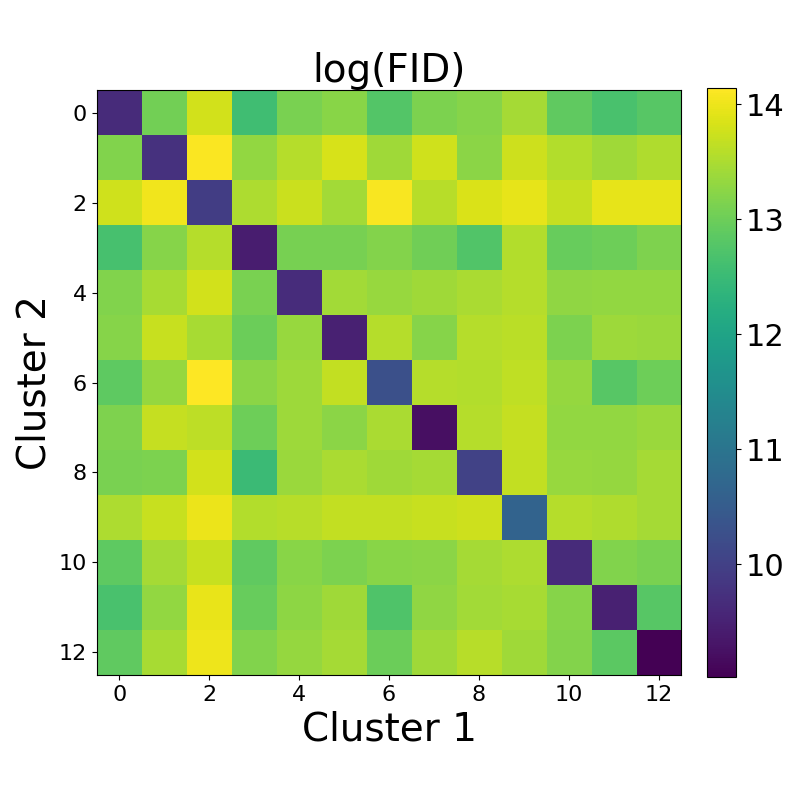}
    \end{subfigure}%
    \begin{subfigure}[t]{0.5\textwidth}
        \centering
        \includegraphics[height=2.5in]{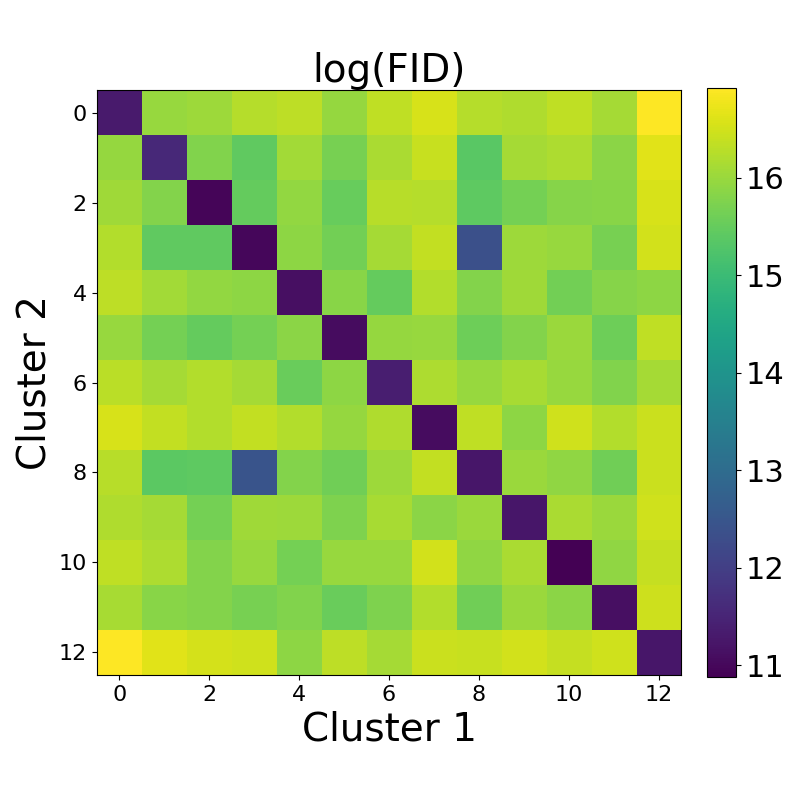}
    \end{subfigure}
    \vspace{-0.8cm}
    \caption{\textbf{FID Between FoldSeek Clusters} \emph{Left:} Pairwise FIDs between FoldSeek clusters computed with GearNet embeddings. Correlation with OT-TMScores is 0.66.  \emph{Right:} Pairwise FIDs between FoldSeek clusters computed with split ESM3 embeddings. Correlation with OT-TMScores is 0.7.}
    \label{fig:app_foldseek_clusters}
    \vspace{-0.5cm}
\end{figure}

\begin{figure}
    \vspace{-0.4cm}
    \centering
    \begin{subfigure}[t]{0.5\textwidth}
        \centering
        \includegraphics[height=1.5in]{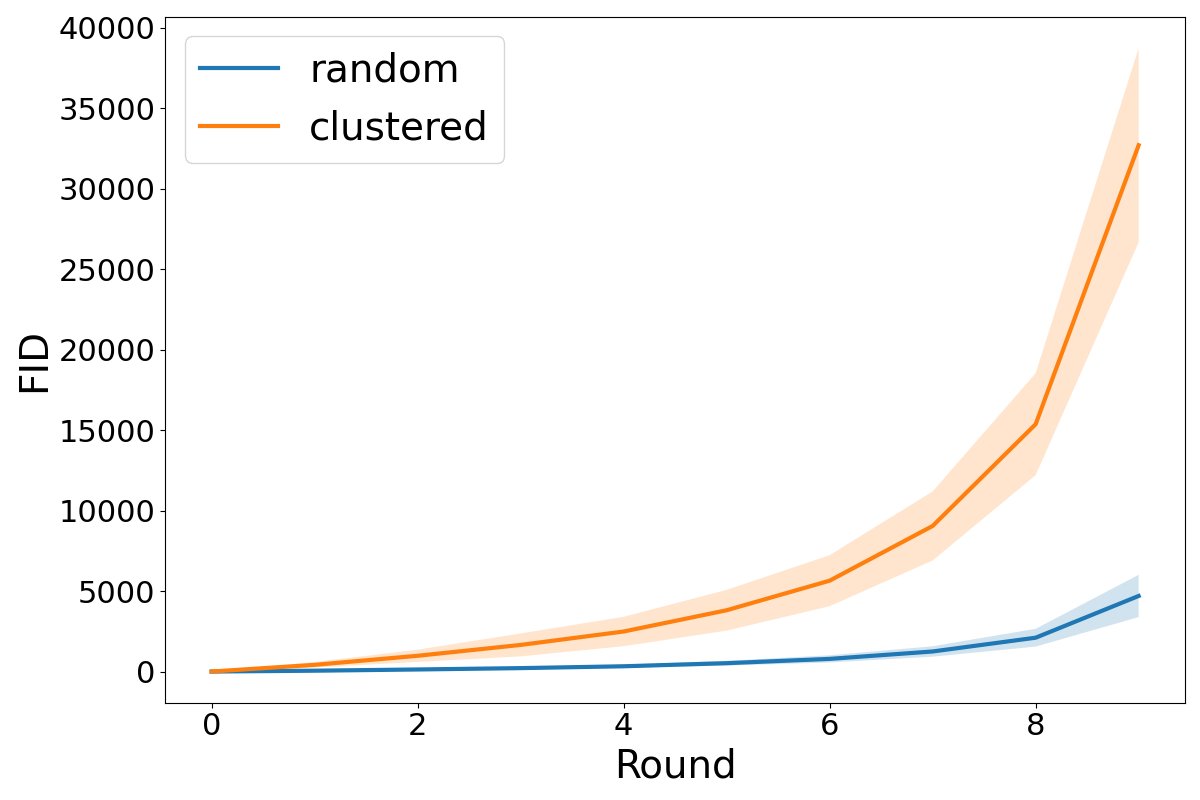}
    \end{subfigure}%
    \begin{subfigure}[t]{0.5\textwidth}
        \centering
        \includegraphics[height=1.5in]{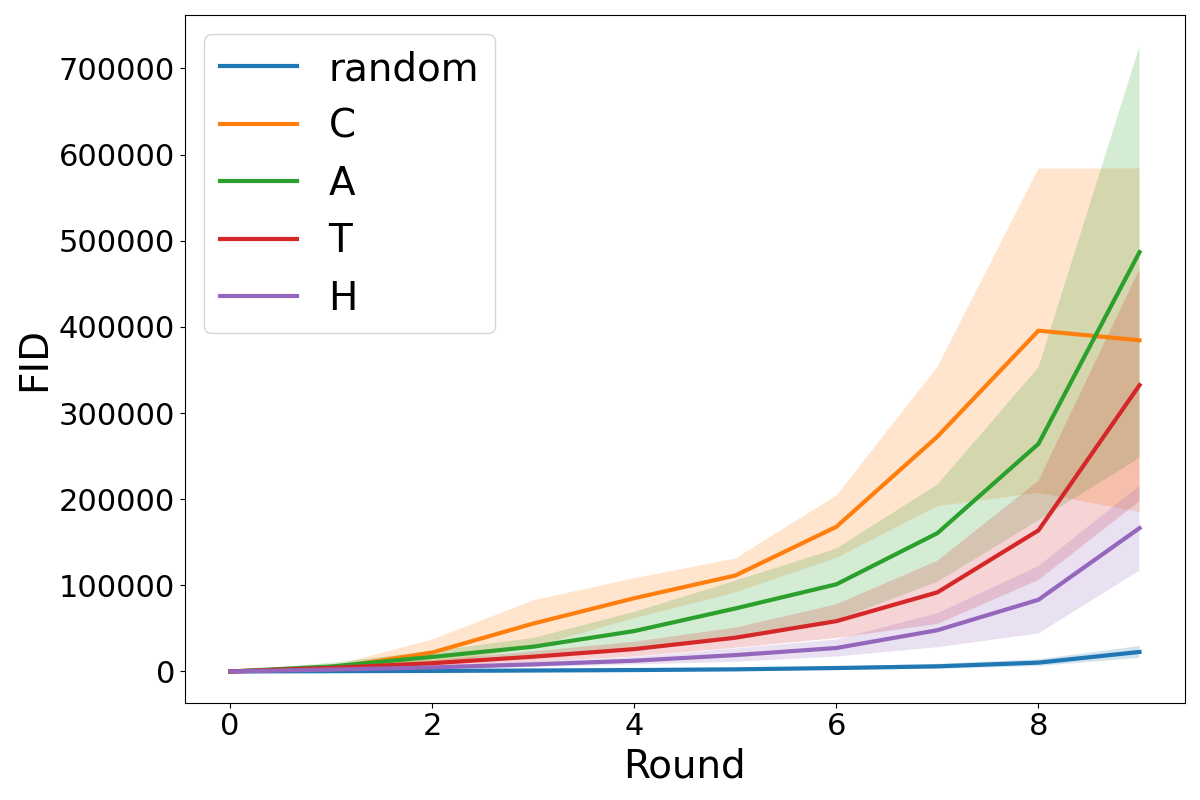}
    \end{subfigure}

    \vspace{0.5cm} 

    \begin{subfigure}[t]{0.5\textwidth}
        \centering
        \includegraphics[height=1.5in]{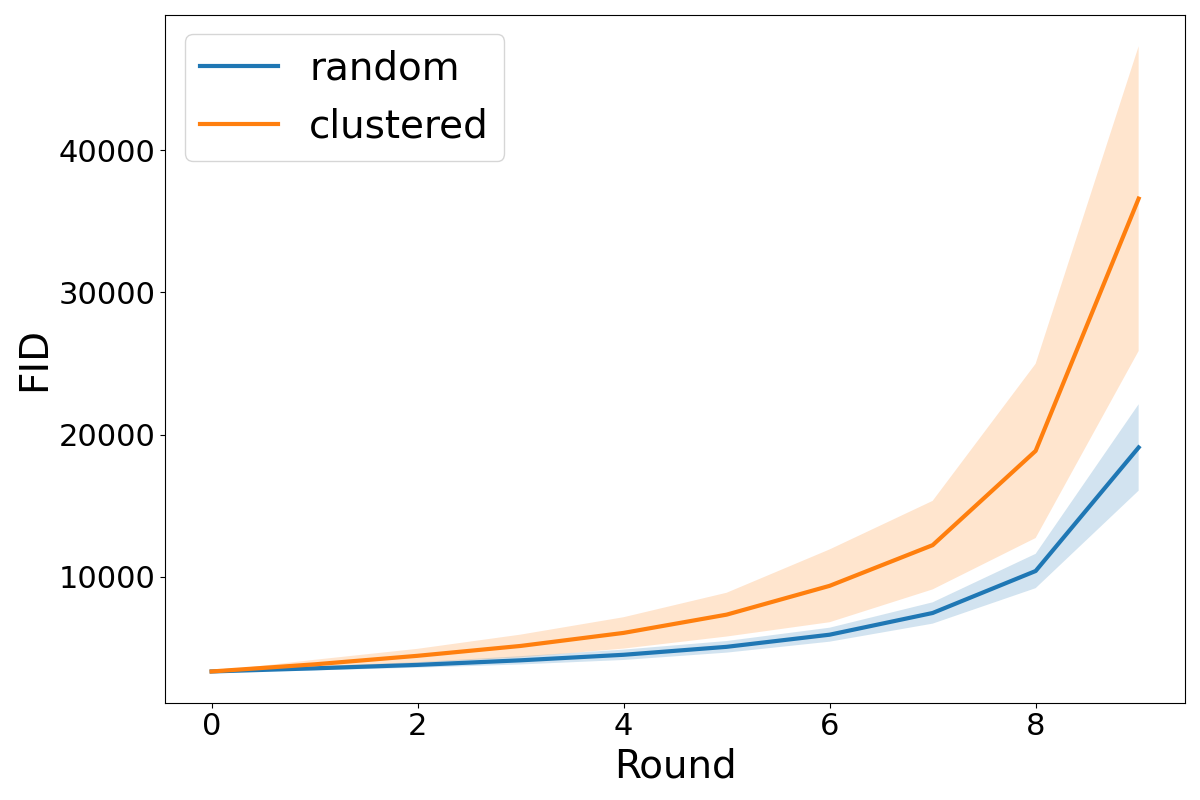}
    \end{subfigure}%
    \begin{subfigure}[t]{0.5\textwidth}
        \centering
        \includegraphics[height=1.5in]{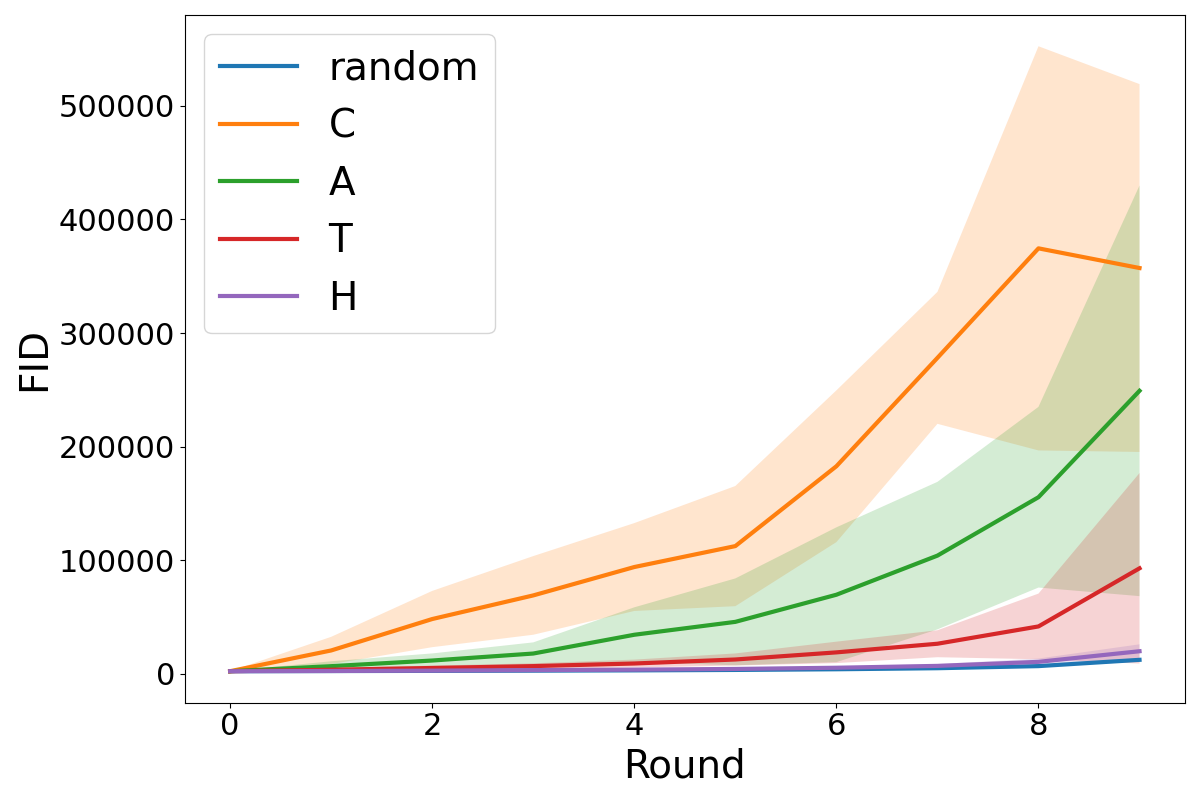}
    \end{subfigure}
    \caption{\textbf{Diversity races.} \textit{Top row}: Diversity races conducted using GearNet embeddings. \textit{Bottom row}: Diversity races conducted using split ESM3 embeddings. \textit{Left}: Race conducted with FoldSeek clusters. \textit{Right}: Race conducted with CATH clusters, with one racer for each level of the hierarchy.}\label{fig:app_diversity-race}
    \vspace{-0.4cm}
\end{figure}

In addition to ESM3, we also explore using GearNet \citep{zhang2022protein} to compute embeddings. As can be seen in Fig.~\ref{fig:cov_ranks}, we also find the need to project the embeddings down to a lower dimension, using the same approach as for ESM3. Using these embeddings to compute FIDs, we report the same experiments as for ESM3, including FIDs of perturbations (Fig.~\ref{fig:app_perturbations}), FIDs between FoldSeek Clusters (Fig.~\ref{fig:app_foldseek_clusters}, and diversity races (Fig~.\ref{fig:app_diversity-race}). Overall, we find that the GearNet-based FID behaves less consistently than the ESM3-based one. For example, adding 0.5\AA{} of random noise to PDB structures gives a lower FID than AF3-refolded structures, the correlation of the inter-cluster FIDs with the OT-TMScores is only 0.66 compared to 0.74 for ESM3, and the order of the CATH clusters is not entirely preserved in the diversty race.

\paragraph{Split ESM3 Embeddings}

ESM3 computes embeddings for each residue in a protein, which we average to compute a whole-protein representation. Instead, we also explore keeping the residue embeddings separate and computing FIDs using the pooled embeddings. This could focus the FID on differences in local structure rather than global structure. Hence, if two sets of structures can be decomposed into similar sets of \textit{local} structures, they would still achieve a low FID even if the \textit{global} structures are different. Because the total number of residues in a set of protein structures can grow very large, we subsample residue embeddings when computing the FIDs. We also still project the embeddings to a lower dimension. Just as for GearNet, we reproduce the same experiments as in the main paper but using these split embeddings. Interestingly, we see that the FID of AF3-refolded samples in Fig.~\ref{fig:app_perturbations} is closer to the PDB samples when compared to using averaged ESM3 embeddings. The ratio of AF3-refolded to PDB FIDs is around 19 when using averaged embeddings, and only around 6 with split embeddings. We hypothesize that, without access to MSAs, AlphaFold3 is still able to reconstruct local structures faithfully, although it may fail to produce accurate global structures. This suggests that the split ESM3 embeddings indeed focus more on local differences than global differences. This may be attractive in some cases, although for the purposes of our experiments we find that overall the averaged embeddings still perform better when looking at the inter-cluster FIDs (Fig.~\ref{fig:app_foldseek_clusters}) and diversity races (Fig.~\ref{fig:app_diversity-race}).

\section{Embedding Dimension for ESM3}

\begin{figure}
    \centering
    \includegraphics[width=\linewidth]{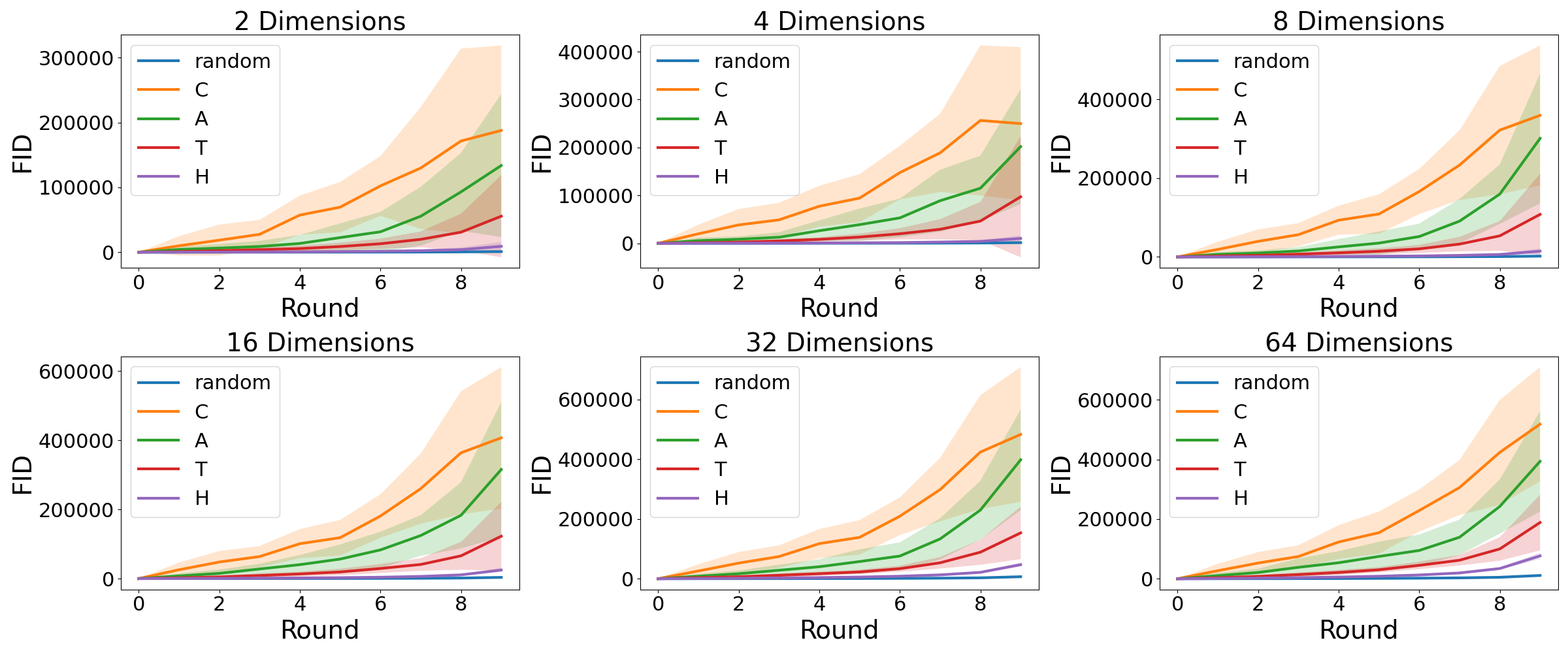}
    \caption{\textbf{Diversity Races with Varying Dimension} CATH diversity races for different embedding dimensions.}
    \label{fig:cath_race_dim}
\end{figure}

\begin{table}
    \centering
    \begin{tabular}{lrrrrrr}
        \toprule
        Dimension: & 2 & 4 & 8 & 16 & 32 & 64 \\
        \midrule
         Correlation: & 0.669 & 0.720 & 0.729 & 0.734 & 0.736 & 0.667 \\
         \bottomrule
    \end{tabular}
    \caption{\textbf{Correlation to OT-TMScore with Varying Dimension} Correlations of FIDs with OT-TMScore for pairwise FoldSeek cluster comparisons.}
    \label{tab:corr_dim}
\end{table}

We found it useful to project the ESM3 embeddings to a lower dimension using PCA. Here we perform a sweep of the embedding dimension, reporting results when using different numbers of embeddings. Fig.~\ref{fig:cath_race_dim} shows the CATH diversity races when run using different embedding dimensions. We see that if we project onto too few dimensions, we start to lose the separation and ordering of the racers. Additionally, Table~\ref{tab:corr_dim} shows the correlation of FID to OT-TMScore across the FoldSeek cluster comparisons from Table.~\ref{fig:foldseek_clusters}. We thus opted to use 32 dimensions since this achieved the best correlation with OT-TMscore, showed good results in the CATH race, and did not lead to any numerical issues when computing FIDs.

\end{appendices}

\end{document}